\documentclass[12pt]{article} 

\usepackage{amsmath,amssymb,graphicx} 
\usepackage{epsf} 
\usepackage{cite} 

\newcommand{\beq}{\begin{eqnarray}}
\newcommand{\eeq}{\end{eqnarray}} 

\newcommand{\centeron}[2]{{\setbox0=\hbox{#1}\setbox1=\hbox{#2}\ifdim 
                           \wd1>\wd0\kern.5\wd1\kern-.5\wd0\fi \copy0 
                           \kern-.5\wd0\kern-.5\wd1\copy1\ifdim\wd0>\wd1 
                           \kern.5\wd0\kern-.5\wd1\fi}} 
\newcommand{\ltap}{\>\centeron{\raise.35ex\hbox{$<$}} 
                   {\lower.65ex\hbox{$\sim$}}\>} 
\newcommand{\gtap}{\>\centeron{\raise.35ex\hbox{$>$}} 
                   {\lower.65ex\hbox{$\sim$}}\>}

\newcommand\ZZ{\hbox{\zfont Z\kern-.4emZ}} 
\font\zfont = cmss10 
\newcommand{\sfrac}[2]{{\textstyle\frac{#1}{#2}}}

\def\tv#1{\vrule height #1pt depth 5pt width 0pt} 
\def\tvbas#1{\vrule height 0pt depth #1pt width 0pt}

\renewcommand{\theequation}{\thesection.\arabic{equation}}

\newcommand{\al}{\alpha}

\newcommand{\x}{\chi}

\newcommand{\pd}{\partial}

\begin{document} 

\begin{titlepage} 

\begin{flushright} 
{\tt hep-ph/0310355} \\ 
Saclay t03/147\\ 
\end{flushright} 

\vspace*{0.5cm} 

\begin{center} 
{\huge \bf Fermions on an Interval:} \\ 
\vspace{.3cm}
{\huge \bf Quark and Lepton Masses}\\
\vspace{.2cm} 
{\huge \bf without a Higgs} 
\end{center} 
\vspace*{0.2cm}

\begin{center} 
{\bf 
{Csaba Cs\'aki}$^{a}$, 
 {Christophe Grojean}$^{b,c}$, 
 {Jay Hubisz}$^{a}$,\\ 
{Yuri Shirman}$^{d}$, 
{\rm and} 
{John Terning}$^{d}$} 
\end{center} 
\vspace*{8pt} 

\begin{center} 
$^{a}$ {\it Institute for High Energy Phenomenology, Laboratory of Elementary Particle Physics, \\ 
Cornell University, Ithaca, NY 14853} \\ 
\vspace*{0.1cm} 
$^{b}$ {\it Service de Physique Th\'eorique, CEA Saclay, F91191 Gif-sur-Yvette, France} \\ 
\vspace*{0.1cm} 
$^{c}$ {\it Michigan Center for Theoretical Physics, 
Ann Arbor, MI 48109, USA}\\ 
\vspace*{0.1cm} 
$^{d}$ {\it Theory Division T-8, Los Alamos National Laboratory, Los Alamos, 
NM 87545} \\ 
\vspace*{0.3cm} {\tt  csaki@mail.lns.cornell.edu, grojean@spht.saclay.cea.fr, 
hubisz@mail.lns.cornell.edu, shirman@lanl.gov, terning@lanl.gov} 
\end{center} 

\vspace*{.3cm} 

\begin{abstract} 

\vskip 3pt \noindent We consider fermions on an extra dimensional interval. We find the 
boundary conditions at the ends of the interval that are consistent with the variational principle,
and explain which ones arise in  various physical circumstances. We apply these results to higgsless 
models of electroweak symmetry breaking, where electroweak symmetry is not broken by a scalar
vacuum expectation value, but rather by the boundary conditions of the gauge fields. We show that it is 
possible to find a set of boundary conditions for bulk fermions that would give a realistic fermion
mass spectrum without the presence of a Higgs scalar, and present some sample  fermion mass spectra
for the standard model quarks and leptons as well as their resonances.

\end{abstract} 
\end{titlepage} 

\newpage 

\section{Introduction} 
\setcounter{equation}{0}  
The most exciting question facing particle physics is how electroweak
symmetry is broken in nature. Since the scattering of massive
$W$ and $Z$ bosons violate unitarity at the scale of $\sim 1.8$ TeV, we know 
that some new particles must appear before those scales are reached to 
unitarize these amplitudes or the theory will be strongly interacting. 
In 4D the only possibility to unitarize these scattering amplitudes (with a 
single particle) is via the exchange of a scalar Higgs particle.
It has been recently pointed out in~\cite{CGMPT} that extra dimensions 
may provide an alternative way for unitarizing the scattering of the massive 
gauge bosons via the exchange of a tower of massive Kaluza-Klein (KK)
gauge bosons (see also~\cite{otherunitarity,SonStephanov}). In this case 
electroweak symmetry would be broken not by the expectation value of a scalar 
(or a scalar condensate), but rather by the boundary conditions (BC's) for the gauge fields.\footnote{For 
other possibilities of utilizing extra dimensions for electroweak symmetry breaking 
see~\cite{A5Higgs,ST}.} As long as the 
BC's are consistent with the variation of a fully gauge invariant action,
the symmetry breaking will be soft in the sense that the UV properties of the 
scattering amplitudes will be as in the higher dimensional gauge theory~\cite{CGMPT}.

A model of higgsless electroweak symmetry breaking (EWSB) with a realistic gauge structure has been presented 
in~\cite{CGPT}. However, the Higgs scalar of the standard model (SM)
serves two purposes: besides breaking the electroweak symmetry it is also necessary for the 
generation of fermion masses without explicitly breaking gauge invariance. The purpose of this paper
is to examine how fermion masses can be generated in higgsless models where EWSB happens via BC's 
in extra dimensions. 

The structure of these higgsless models is generically of the following form~\cite{CGPT}: we consider the 
a modification of the Randall-Sundrum model~\cite{RS} with gauge fields in the bulk~\cite{RSbulk,CET},
where the bulk gauge group is SU(2)$_L\times$SU(2)$_R\times$U(1)$_{B-L}$. The addition of the second
SU(2)$_R$ in the bulk is necessary~\cite{ADMS} in order to ensure the presence of a custodial SU(2)$_R$ symmetry 
in the holographic interpretation~\cite{holography}. On the Planck brane SU(2)$_R\times$ U(1)$_{B-L}$ 
is broken to U(1)$_Y$, while EWSB happens on the TeV brane where SU(2)$_L\times$SU(2)$_R \to$SU(2)$_D$
(an early flat space version of this model was presented in~\cite{CGMPT}, and recently re-examined 
in~\cite{BPR}). Two important issues regarding the warped higgsless model that were not addressed fully in~\cite{CGPT}
were the generation of fermion masses without a Higgs, and the corrections to electroweak precision observables.
The issue of fermion masses will be discussed in detail in this paper. There are several potential sources 
for corrections to electroweak precision observables in a higgsless model:
the enlarged gauge structure, the missing Higgs scalar, and the modified fermion sector. Recently
\cite{BPR} examined the $S$ parameter in the flat space version of this model and found that analogously
to technicolor theories there is a large positive contribution. However, no comprehensive analysis of 
the electroweak observables including all sources of corrections listed above has been done to 
date. We plan to address these issues for the case of an AdS$_5$ bulk (as considered in \cite{CGPT,nomura}) in a forthcoming publication.

In order to be able to generate a viable spectrum and coupling for the SM fermions, the fermions have
to feel the effect of EWSB, so they need to be connected with the TeV brane. However they can't be simply 
put on the TeV brane, since in that case they would form multiplets of SU(2)$_D$, which they don't. 
Thus the fermions also have to be put into the bulk, as 
in~\cite{matthias,GherPom,kaptait,HS1,nomurasmith,HSrecent}. We assume that the left handed SM fermions
will form SU(2)$_L$ doublets and the right handed ones SU(2)$_R$ doublets (including a right handed neutrino).
Since 5D bulk fermions contain two 4D Weyl 
spinors (like a 4D Dirac fermion), one has to first make sure that in every 5D bulk fermion there is only 
a single 4D Weyl spinor zero mode. These zero modes will be identified with the usual SM fermions. 
In order to recover the usual gauge coupling 
structure for the light fermions, the zero modes for the light fermions 
 have to be localized close to the Planck brane. Since the theory on the TeV brane is vector-like, one can simply 
add a mass term on the TeV brane that connects the left and right handed fermions. This is however not sufficient,
since the gauge group on the TeV brane would force the up-type and down-type fermions to be degenerate. 
The splitting between these fermions can be achieved by mixing the right handed fermions
with fermions localized on the Planck brane where 
SU(2)$_R$ is broken (which is equivalent to adding different Planck-brane induced kinetic terms for the 
right handed fermions). The detailed models for the fermion masses for the warped space higgsless model will be
presented in Section 7 (while the analog constructions for the somewhat simpler flat-space toy model can be
found in Section 5). 

Before we discuss fermion mass generation for the higgsless models in detail, we will discuss 
the general issues surrounding the often confusing subject of BC's and masses for fermions in one 
extra dimension. In Section 2 we examine the possible BC's for fermions on an interval that are consistent with 
the vanishing of the boundary variations of the action. This is an extension of the general discussion of
\cite{CGMPT} of BC's in an extra dimension to the fermion sector. In Section 3 we discuss the general KK 
decomposition for fermions on an interval (with some simple
examples worked out in detail in Appendix B), while in Section 4 we give the physical interpretation of the 
various BC's obtained from the variational principle. More important examples for BC's in the presence of 
mixing of bulk fermions on a brane are presented in Appendix C. In Section 5 we apply the results of Sections 2-4
to propose the BC's for the fermion sector of the flat space higgsless model. In Section 6 we discuss the 
general issues of fermionic BC's in warped space, and then finally present the BC's and mass spectra for the
warped space higgsless model in Section 7. We conclude in Section 8. Appendix A contains notations and
spinor conventions.  

\section{Fermion Boundary Conditions from the Variational Principle} 
\label{sec:BCs}
\setcounter{equation}{0}  

We start by considering a  theory of $5D$ fermions on an interval of length $L$, 
with a bulk Dirac mass $m$, and possibly also 
masses for the component  fermions on the boundaries. 
For the moment we will assume that the geometry of the interval is flat and we will come back later to the phenomenologically more interesting case of an AdS$_5$ interval. In 5D, the 
smallest irreducible representation of the Lorentz group is the Dirac spinor, which of course 
contains two two-component spinors from the 4D point of view.  The bulk action for 
the Dirac spinor $\Psi$ is given by the usual form 
\begin{equation} 
        \label{eq:BulkAction}
S = \int d^5 x 
\left( \frac{i}{2}
( \bar{\Psi}\, \Gamma^M \partial_M \Psi
- \partial_M \bar{\Psi}\,  \Gamma^M  \Psi )
- m \bar{\Psi} \Psi  \right)
\end{equation} 
where $M=0,1,2,3,5$. Usually, in the second term the differential operator is integrated by parts
and gives a contribution identical to the first term. However when the fifth dimension is a finite interval, boundary terms appear in the process of  integrating by parts
and using the conventional form of the action would require us to explicitly  introduce those boundary terms. That this is the convenient starting point  can be seen from the 
fact that the action is real or equivalently that the corresponding Hamiltonian is 
hermitian.Writing out the action in terms of the two-component spinors contained in the 5D Dirac spinor as 
\begin{equation} 
\Psi = \left( 
\begin{array}{c} \chi_\alpha 
\\ 
\tv{13}
\bar{\psi}^{\dot{\alpha}} 
\end{array} 
\right) 
\label{bigpsi}
\end{equation} 
and integrating by parts in the 4D coordinates where we do require that the fields vanish at 
large distances, we obtain the following Lagrangian for the two-component spinors (for 
spinor and gamma matrices conventions, see Appendix~\ref{app:conventions}): 
\begin{equation} 
S = \int d^5 x 
\left( 
- i \bar{\chi} \bar{\sigma}^{\mu} \partial_\mu \chi 
- i \psi \sigma^{\mu} \partial_\mu \bar{\psi}  
+\sfrac{1}{2}\, ( \psi  \overleftrightarrow{\partial_5}  \chi   
-  \bar{\chi}  \overleftrightarrow{\partial_5}   \bar{\psi}   )
+ m (\psi \chi +  \bar{\chi} \bar{\psi} )
\right),
\end{equation} 
where $\overleftrightarrow{\partial_5} = \overrightarrow{\partial_5}-\overleftarrow{\partial_5}$, with the arrows indicating the direction of action of the differential operator.
Varying the action with respect to $\bar{\chi}$ and $\psi$ we obtain the 
standard bulk equations of motion which are given by 
\begin{eqnarray} 
-i \bar{\sigma}^{\mu} \partial_\mu \chi - \partial_5 \bar{\psi} + m \bar{\psi} = 0, 
\nonumber \\ 
-i \sigma^{\mu} \partial_\mu \bar{\psi} + \partial_5 \chi + m  \chi = 0. 
\end{eqnarray} 
However one needs to be careful with the variation, since one needs to do an integration 
by parts in the extra dimension, which will give an extra term for the variation of the 
action on the 
boundaries of the interval. Requiring that the boundary term in the variation vanishes  will give the desired boundary conditions for the fermion fields (we denote by $[X]_0^L$ the quantity 
$X_{|L}-X_{|0}$): 
\begin{equation} 
        \label{eq:BCnomass} 
\delta S_{bound}= \frac{1}{2} \int d^4x 
\left[  
\delta \x\, \psi 
- \delta \psi\, \chi 
- \delta {\bar \psi}\, {\bar \chi} 
+ \delta {\bar \chi}\, {\bar \psi}  
\, \right]_0^L = 0. 
\end{equation} 

To proceed further we need to specify the boundary conditions. These have to be such that 
the boundary variation in (\ref{eq:BCnomass}) vanishes. Note, that this is a somewhat unusual 
boundary variation term (at least compared to the case of scalar and gauge fields) since it mixes 
the two Weyl spinors. We will first discuss the simplest and most commonly adopted solutions, 
and then consider the more general cases. The most obvious solution to enforce the 
vanishing of (\ref{eq:BCnomass}) is by fixing one of the two spinors to zero on the endpoints, 
for example 
\begin{equation} 
\psi _{| 0,L}=0. 
\end{equation} 
As soon as we imposed this condition, we also have that $\delta \psi_{| 0,L}=0$, and the full 
boundary variation term vanishes. This would naively suggest that $\chi$ remains arbitrary at 
the endpoints, however this is not the case, since we still have to require that the bulk 
equations of motion are satisfied everywhere, including at the endpoints of the interval. Since 
the bulk equations mix $\psi$ and $\chi$, when $\psi =0$ we get a first order equation for just 
$\chi$, which can be considered as the boundary condition for the $\chi$ field: 
\begin{equation} 
\left( 
\partial_5 \chi + m \chi 
\right)_{|0,L}=0. 
\end{equation} 
In the limit of $m\to 0$ this is the BC that is usually employed when considering orbifold 
projections. The usual argument is that if one assigns a definite parity to $\chi$ and $\psi$ 
under $y\to -y$, then due to the term $\psi \partial_5 \chi$ in the bulk, 
$\chi$ and $\psi$ have to have opposite parities, so if $\psi$ is chosen to have negative parity 
(that is it vanishes on the endpoints) then $\chi$ has to be positive, so its derivative should vanish. 
This is basically what we see in this simplest solution, except that it is very easy to deal with the 
bulk mass term. If one were to think of this interval as the orbifold projection of a circle, then the 
only way one can fit a bulk mass into the picture is if the bulk mass 
is assumed to switch signs at the orbifold fixed points (the mass 
itself has negative parity), which then makes figuring out the right 
BC's in the presence of the mass term quite hard. We can see that in the interval formulation one does 
not have to worry about such subtleties. 

We have seen above that the simplest possible solutions to the vanishing of the boundary 
variation leads to the boundary conditions generically employed when considering orbifold constructions.  
However, one does not need to require the individual terms in 
(\ref{eq:BCnomass})  to vanish, it is sufficient for the whole sum to 
vanish.  In fact, requiring the individual variations to vanish 
over-constrains the system, as is clear from a simple counting of the 
degrees of freedom of the theory.  There are two constants associated 
with the solutions of two first order differential equations.  One 
boundary condition at each end of the interval then specifies the 
system.  If one forces the individual terms in the boundary variation 
on one endpoint to vanish, then there is no freedom of boundary 
conditions on the opposite endpoint.  Thus one should generically only 
impose one BC at each end of the interval.  Such a BC expresses one of 
the spinors in terms of the other. 
With that in mind 
we can see that the most general solution to the vanishing of the boundary variation is 
when, on the boundary, the two fields $\psi$ and $\chi$ are proportional to each other: 
\begin{equation} 
\label{bc1} 
\psi_{\alpha} {}_{|0,L} = 
\left( 
M_\alpha{}^\beta\, \chi_\beta + N_{\alpha {\dot \beta}}\, \bar{\chi}^{\dot \beta} 
\right)_{|0,L} 
\end{equation} 
where $M$ and $N$ are two matrices that may involve some derivatives 
along the dimensions of the boundary. The action will then have a vanishing 
boundary variation provided that $M$ and $N$ satisfy the two conditions 
\begin{equation} 
M_\al{}^\beta = \sigma_M \epsilon^{\beta\gamma} M_\gamma{}^\delta \epsilon_{\delta\al} 
\ \ {\rm and}\ \ \ 
N_{\al {\dot \beta}} = \sigma_N N^\dagger_{\al {\dot \beta}}  
\end{equation} 
where $\sigma_{M,N}$ are signs due to some possible integration by part of the differential operators contained in $M$ and $N$. Two simple solutions are 
\begin{eqnarray} 
& \displaystyle 
M_\al{}^\beta = c\, \delta_{\al}{}^\beta \ \ \ \mathrm{for\ any\ constant\ } c,
\\ 
& \displaystyle 
N_{\al {\dot \beta}} = i c\, \sigma^\mu_{\al {\dot \beta}} \, \partial_\mu\ \ \ 
\mathrm{for\ any\ real\ constant\ } c. 
\label{bc2} 
\end{eqnarray} 
Note that for fermions belonging to a complex representation of the 
gauge group, gauge invariance requires that either the operator $M$ or its 
inverse vanishes. 
Let us discuss in more detail the solutions of the type 
$\psi_{|0,L}= C_{0,L}\, \chi_{|0,L}$, for arbitrary values of $C_{0,L}$.  A better way of 
expressing this condition is by saying that some linear combination of the two fermion 
fields has to satisfy a Dirichlet boundary condition on both ends. However, these can be 
different combinations on the two sides: 
\begin{equation} 
        \label{eq:angleBC} 
s_{0,L}\, \psi_{|0,L} + c_{0,L}\, \chi_{|0,L}  =  0 
\end{equation} 
where $s_{0,L}$  ($c_{0,L}$) stand for the sine (cosine) of some (possibly complex) angles, $\alpha_{0,L}$, that determine which linear combination of the fields on the two boundaries are vanishing. If there 
are gauge symmetries in the bulk, under which the fermions transform, then $\psi$ and 
$\chi$ transform in complex conjugate representations, as can be seen from Eq. (\ref{bigpsi}). This means, that it is only possible to mix the two fields on the boundary with non-trivial angles if the fermion is in a real 
representation. Thus for real representations $s_{0,L}$ could in principle be arbitrary, however 
for complex representations the only possibilities are $s_{0,L}=0$ or  $s_{0,L}=1$. We will see later on that this choice
of BC's corresponds to adding a Majorana mass on the brane. 

\section{Kaluza-Klein Decomposition} 
\label{sec:KKDecomp} 
\setcounter{equation}{0} 

Now we would like to discuss how to perform the Kaluza--Klein decomposition of these fields. 
In general, when the fermion belongs to a complex representation of the symmetry group, the KK modes can only acquire Dirac masses and the KK decomposition is of the form
\begin{eqnarray} 
        \label{eq:DiracKK}
\chi  = \sum_n g_n(y)\, \chi_{n} (x), \\ 
\bar{\psi} = \sum_n f_n(y)\, \bar{\psi}_n (x), 
\end{eqnarray} 
where $\chi_n$ and $\psi_n$ are 4D two-component  spinors which form a Dirac spinor of mass 
$m_n$ and satisfy the 4D Dirac equation:
\begin{eqnarray} 
-i \bar{\sigma}^{\mu} \partial_\mu \chi_{n} + m_n\, \bar{\psi}_n = 0, \\ 
-i \sigma^{\mu} \partial_\mu \bar{\psi}_n + m_n\, \chi_{n} = 0. 
\end{eqnarray} 
Plugging this expansion into the bulk equations  we get the following set of coupled first order differential equations for the wave functions $f_n$ and $g_n$: 
\begin{eqnarray} 
        \label{eq:1stOrder1}
g_n' + m\, g_n - m_n\, f_n = 0, \\
        \label{eq:1stOrder2}
f_n' - m\, f_n + m_n\, g_n = 0. 
\end{eqnarray} 
Combining the two equations we get as usual decoupled second order equations: 
\begin{eqnarray} 
g''+(m_n^2-m^2)g = 0,\\ 
 f''+(m_n^2-m^2)f= 0. 
\end{eqnarray} 
Depending on the sign of $m_n^2-m^2$ the wave functions $g_n$ and $f_n$ will be either sines and 
cosines or sinhes and coshes (we define $\mathrm{ccos\,} k_nL = \cosh k_nL$ for 
$k_n^2=m^2-m_n^2>0$ 
and $\mathrm{ccos\,} k_nL = \cos k_nL$ for $k_n^2=m_n^2-m^2>0$ and similarly for 
$\mathrm{ssin\,} k_nL$): 
\begin{eqnarray} 
        \label{eq:wv1}
g_n(y) = A_n \, \mathrm{ccos\,} k_n y + B_n\,  \mathrm{ssin\,}  k_n y,\\ 
        \label{eq:wv2}
f_n (y) = C_n \, \mathrm{ccos\,}  k_n  y + D_n \, \mathrm{ssin\,} k_n y. 
\end{eqnarray} 
These wave functions are analogous to the ones obtained for bosonic fields. For fermions, the bulk equations are going to teach us something more about the wave functions. Indeed the first order
coupled differential equations (\ref{eq:1stOrder1})-(\ref{eq:1stOrder2}) relate the coefficients $A_n, B_n, C_n, D_n$ to each other. Using the form (\ref{eq:wv1})-(\ref{eq:wv2}) of the wave functions and for $m_n \neq 0$, the two bulk equations are equivalent to one another and impose, for $k_n^2=m_n^2-m^2>0$,
\begin{eqnarray} 
m C_n -  k_n D_n - m_n A_n =0, \\ 
k_n C_n + m D_n - m_n B_n =0. 
\end{eqnarray} 
When $m^2-m_n^2>0$, the sign of the term involving $k_n$ in the second equation is flipped.

The boundary conditions may also allow the presence of a zero mode which can have a non trivial profile of the form (\ref{eq:wv1})-(\ref{eq:wv2}) with $k_n^2=m^2$ for a non vanishing bulk mass.
For the case of the zero mode the bulk equations (\ref{eq:1stOrder1})-(\ref{eq:1stOrder1}) are decoupled
and simply reduce to
\begin{equation} 
A_0 = -B_0 \ \ \rm{and} \ \ \ C_0 = D_0. 
\end{equation} 

Some explicit examples of KK decomposition are given in Appendix~\ref{app:KKdecomposition} when BC's of the form 
$s \psi_| + c \chi_|$ are imposed at $0^+$ and $L^-$. We also discuss there how to amend the form of the decomposition (\ref{eq:DiracKK}) when the gauge quantum numbers of the fermion allow the KK modes
to have Majorana masses.

Before we close this section, we just remind the reader what the status of all of these various 
boundary conditions with respect to each other is: that is can we have the different modes 
corresponding to {\it different} boundary conditions present in the theory at the same time? 
The answer is no. A theory is obtained by fixing the BC's for the fields (picking one of the possibilities from the list given above) 
once and for all. If we were to include modes corresponding to different BC's into the theory, we would loose hermiticity of the Hamiltonian, that is the theory would no longer be unitary. If one were to insist on putting these different modes together, an 
additional super-selection rule would have to be added that would make these different modes orthogonal, signaling that they belong to a different sector of the Hilbert space, which practically means that a new quantum number corresponding to the choice of 
the BC would have to be added. But usually this is avoided by simply considering only a fixed BC.

\section{Physical Interpretation of the Boundary Conditions}
\setcounter{equation}{0}
\label{sec:dynbc} 

We would like to have an intuitive physical picture of the various
fermionic boundary conditions.  Unlike a scalar field, which in the
absence of boundary interactions naturally has a flat profile
($\partial_5 \phi=0$) on the boundary, the fermions cannot have a purely
flat wave function.  This is a result of the dynamics of 5D fermions, which can
be broken up into two two-component spinors.  The bulk equations
of motion for these spinors imply that, in the absence of a bulk Dirac
mass (as we will assume throughout this section), if one spinor profile has zero derivative at a boundary, then
the opposite spinor must obey Dirichlet boundary conditions, and vice
versa. However, we have seen in Section 2 that there is a variety of BC's that one can impose instead of the
simplest $\partial_5 \chi_| =\psi_| =0$ condition. The purpose of this section (and its
continuation in Appendix C) is to understand what physical situations the various  BC's
(some of which may seem quite obscure at first sight) correspond to. 

What we would like is to be able to consider a setup with arbitrary localized mas\-ses/mix\-ings/ki\-ne\-tic terms
and translate these into some BC's similar to the form of (\ref{bc1}). However, it is not easy to arrive at these
BC's from the variational principle if the localized terms are directly added at the boundary. The reason is that
due to the first order nature of the bulk equations of motion the presence of a localized term necessarily implies 
a discontinuity in some of the wave functions. If the localized term is added directly at the boundary, one would
have to treat the values and variations of the fields at the boundaries as independent from the bulk values which 
makes the procedure very hard to complete. Instead, our general approach to treating the localized terms will 
be the following:
\begin{itemize}
\item Push the localized terms at a distance $\epsilon$ away from the boundary, which implies
the presence of a $\delta$-function in the bulk equations of motion;
\item Impose the simplest BC's $\partial_5 \chi_| =\psi_| =0$ at the real boundary $y=0,L$;
\item By combining the jump equation at $y=\epsilon$ with the BC's at $y=0$ obtain a relation between the fields
at $y=\epsilon$;
\item Take the limit $\epsilon \to 0$ and treat the relation among the fields at $y=\epsilon$ as the BC's for the 
theory on an interval with the localized terms added on the boundaries. 
\end{itemize}
By construction, the BC's obtained this way will always satisfy the variational principle (that is make 
(\ref{eq:BCnomass}) or its analog in the presence of more fields vanish), 
but this way the physical interpretation of the 
possible parameters appearing in the BC's will become clear. 

In this section we will first show what a possible physical realization of the usually applied simple
BC $\partial_5 \chi_| =\psi_| =0$ is, and then consider adding a Majorana mass on the boundary as discussed above. 
Many more important examples of adding localized terms will be discussed in Appendix C.

\subsection{Physical Realization of the Simplest Dirichlet--Neumann BC's} 

The easiest way to realize the simple $\partial_5 \chi_| =\psi_| =0$  boundary condition is to
give the fermions a mass that is a function of an extra dimensional
coordinate, where this extra dimension is infinite in extent.  For
example, one can construct a square well mass term given by
\begin{equation}
m( y ) = m_- \,\theta (-y) + m_+\, \theta (y-L)
\end{equation}
where $\theta$ is the usual Heaviside function.
Taking $m_-$ and $m_+$ to be large, yet finite, the solution consists of modes
which tail off exponentially outside of the well.  The wavefunctions of $\chi$ and $\psi$
in region I ($y \leq 0$) and II ($y  \geq L$) are given by
\begin{eqnarray}
(I) \
\left\{ \begin{array}{l}
f_n = C_{n}^-\, e^{k_{n}^-y}\\
\tvbas{5} g_n = A_{n}^-\, e^{k_{n}^-y}
\end{array} \right.
\ \
(II) \
\left\{ \begin{array}{l}
f_n = C_{n}^+ \,e^{k_{n}^+y}\\
\tvbas{5} g_n = A_{n}^+\,  e^{k_{n}^+y}
\end{array} \right.
\end{eqnarray}
where $k_{n}^{\pm\, 2}=m_\pm^2-|m_n^2|$.  As $m$ grows, the exponentials drop
off more and more quickly, and thus the fermions are confined to a ``fat
brane'' of width $L$ (which corresponds to the well). The wavefunctions at $y=0$ and $y=L$ are continuously matched
with the solution within the ``well''
\begin{equation}
(0 \leq y \leq L) \
\left\{ \begin{array}{l}
f_n = C_{n} \, \cos m_n y + D_{n} \, \sin m_n y\\
\tvbas{5} 
g_n = A_{n}\, \cos m_n y + B_{n} \, \sin m_n y.
\end{array} \right. 
\end{equation} 

The existence of a 4D massless mode  depends on the details of the mass profile $m(y)$.
For instance, when $m_+=m_-=m$,
a quick calculation shows that this particular mass background does
not lead to a normalizable chiral zero mode profile. Indeed the ``bulk'' equations of motion
together with the continuity conditions at $y=0$ and $y=L$ leads to the
quantization equation
\begin{equation} m_n \tan m_n L = \sqrt{m^2-m_n^2}, \end{equation} 
which obviously does not allow a massless mode.

However, if one changes the mass profile to $m_+=-m_-=m$, the quantization
equation becomes
\begin{equation} \sqrt{m^2-m_n^2}\, \tan m_n L = - m_n, \end{equation} 
which now supports a massless solution. This is a stepwise
analogue of the well known domain wall localization of chiral
fermions~\cite{Kaplan}.  When $m \rightarrow \infty$, the solution within the well can be
equivalently obtained by ignoring the exterior regions and by imposing
the following boundary conditions (for $m>0$)
\begin{equation}
        \label{eq:basicBC}
\partial_5 \chi_{|0} = 0, \ \
\psi_{|0} = 0, \ \
\partial_5 \chi_{|L} = 0, \ \
\psi_{|L} = 0.
\end{equation}
%

\subsection{BC's in the presence of a brane localized Majorana mass}
\label{sec:FlatBraneMajoranaMass}

We now consider adding localized terms to the fermion action, and 
ask how the simple BC derived above will change in the presence of these terms. 
We will illustrate in detail how to implement the steps outlined at the beginning of this 
section for the case  when a Majorana mass  is added on the boundary. This example in the context of orbifolds
has been discussed in~\cite{BFZ} (see also~\cite{otherBagger}). Several other physical examples
are worked out in Appendix~\ref{app:otherBCs}. 

To be able to add a Majorana mass at the $y=0$ boundary, we need of course to consider
a fermion that belongs to a real representation of the unbroken gauge group. On top of the bulk 
action~(\ref{eq:BulkAction}), we then consider the following brane action (slightly separated from the boundary):
\begin{equation}
S_{4D}
=  
\int d^4 x   \,
 \sfrac{1}{2}\, L \left( M\, \chi \chi  + M^*\, \bar{\chi} \bar{\chi} \right)_{|y=\epsilon}.
\end{equation}
Note that the mass has been written as $ML$ to give to $M$ a mass dimension equal to one.

To find the modified BC's in the presence of this brane mass term, the first step is to chose the 
Dirichlet-Neumann BC's the two two-component spinors
$\chi$ and $\psi$ would have satisfied at the ``real boundary'' $y=0$ 
in the absence of the Majorana mass term.
Thus as previously we assume that:
\begin{equation}
        \label{eq:At0}
\partial_5 \chi_{|0} = 0, \ \
\psi_{|0} = 0, \ \
\partial_5 \chi_{|L} = 0, \ \
\psi_{|L} = 0.
\end{equation}
The effect of the brane mass term is to introduce discontinuities in the wave functions at $y=\epsilon$.
The bulk equations of motion are modified to 
\begin{eqnarray}
& \displaystyle
- i \bar{\sigma}^{\mu} \partial_\mu \chi 
-  \partial_5 \bar{\psi} 
+ m \bar{\psi} 
+ M^*L \, \delta ( y - \epsilon )\, \bar{\chi} 
= 0, 
\\
& \displaystyle
-i \sigma^{\mu} \partial_\mu \bar{\psi} 
+ \partial_5 \chi 
+ m  \chi
 = 0.
\end{eqnarray}
Integrating the first equation over the delta function term shows that,
while $\chi$ remains continuous, the value of the $\psi$ profile undergoes a jump:
\begin{equation}
[ \bar{\psi} ]_{|\epsilon} = M^*L\,  \bar{\chi}{}_{|\epsilon}.
\end{equation}
Because $\psi$ undergoes a jump, the second bulk equation of motion
requires that the derivative of $\chi$ also
undergoes a jump:
\begin{equation}
[ \partial_5 \chi ]_{|\epsilon} = i \sigma^\mu \pd_\mu [ \bar{\psi} ] {}_{|\epsilon}.
\end{equation}
In the limit $\epsilon \to 0$ and from the fixed values (\ref{eq:At0}) of $\chi$ and
$\psi$ at $y=0$, the jump equations finally give the BC at $y=0^+$. This is what we will be 
interpreting as the BC corresponding to the theory with Majorana masses on the boundary $y=0$
(the BC at $y=L$ remain of course unaffected by the mass term localized at $y=0$):
\begin{eqnarray}
        \label{eq:BCMaj}
\partial_5 \chi {}_{|0^+} = i \sigma^\mu \pd_\mu \bar{\psi} {}_{|0^+}-m\chi {}_{|0^+}, \ \
 \psi{}_{|0^+}  =  ML \,  \chi {}_{|0^+},\ \
 \partial_5 \chi {}_{|L^-} = 0, \ \
\psi {}_{|L^-} = 0.
\end{eqnarray}
The first equation is just the bulk equation of motion evaluated at the boundary, so we conclude that the BC
corresponding to the Majorana mass is just
\begin{equation}
\psi{}_{|0^+}  =  ML \,  \chi {}_{|0^+}.
\end{equation}
What one should recognize at this point is that (\ref{eq:BCMaj})
corresponds precisely to the boundary condition
\begin{equation}
(c_0 \psi + s_0 \chi)_{|0^+} = 0.
\end{equation}
with $ML = -s_0/c_0$. Thus we have a completely
dynamical description of one of the boundary conditions mentioned in 
Section~\ref{sec:BCs}. One can also reproduce the other types of BC's by adding different
boundary localized operators like a localized kinetic term or a mass interaction term
with boundary localized fermions. An exhaustive list of cases are worked out
in Appendix~\ref{app:otherBCs}.

\section{Fermion Masses in the SU(2)$_L\times$SU(2)$_R\times$U(1)$_{B-L}$ Flat Space Toy Model}
\label{sec:flat}
\setcounter{equation}{0} 

As an application of the previous sections, we will consider generating masses for leptons 
in a higgsless extra dimensional model in flat space where the bulk gauge group\footnote{With quarks in the bulk there is also a bulk $SU(3)_{\rm color}$ gauge group.} is 
SU(2)$_L\times$ SU(2)$_R\times$ U(1)$_{B-L}$ as presented in Ref.~\cite{CGMPT}. This model has 
large corrections to the electroweak precision observables, and can not be viewed as a realistic model 
for electroweak symmetry breaking. However, most of the large corrections can be eliminated by 
putting the same model into warped space~\cite{CGPT}, or as pointed out recently in~\cite{BPR} by adding 
large brane localized gauge kinetic terms in the flat space case. 
We find it useful to first present 
some of the features of the construction for the fermion masses in the flat space model, as a 
preparation for the more complicated warped case presented at the end of this paper. 

\subsection{Lepton sector}

As always in a left--right symmetric model,  the fermions are in the representations 
$(2,1,-1/2)$ and $(1,2,-1/2)$ of SU(2)$_L\times$ SU(2)$_R\times$ U(1)$_{B-L}$ for left and right handed 
leptons respectively. Since we assume that the fermions live in the bulk, both of these 
are Dirac fermions, thus every chiral SM fermion is doubled (and the right handed neutrino is added 
similarly). Thus the left handed doublet can be written as 
\begin{equation} 
        \label{leftlep}
\left( 
\chi_{\nu_L},
\bar{\psi}_{\nu_L},
\chi_{e_L},
\bar{\psi}_{e_L}
\right)^t
\, ,
\end{equation} 
where $(\chi_{\nu_L},\chi_{e_L})$ will eventually correspond to the SM SU(2)$_L$ doublet
and $({\psi}_{\nu_L},{\psi}_{e_L})$ is its SU(2)$_L$ antidoublet partner needed to form a complete
5D Dirac spinor. Similarly, the content of the right handed doublet is 
\begin{equation} 
\label{rightlep}
\left(
\chi_{\nu_R}, \bar{\psi}_{\nu_R}, \chi_{e_R}, \bar{\psi}_{e_R}
\right)^t\, ,
\end{equation} 
where $(\psi_{\nu_R},\psi_{e_R})$ would correspond to the 'SM' right-handed doublet, i.e., the right
electron and the extra right neutrino, while $(\chi_{\nu_R},\chi_{e_R})$ is its antidoublet partner
again needed to form a complete 5D Dirac spinor.

For simplicity we assume  for now the absence of a bulk Dirac mass term 
(we will later need to introduce such 
terms in the warped scenario to build a realistic model).
In the absence of any brane induced mass terms the fields 
$\chi_{\nu_L}, \chi_{e_L}, \psi_{\nu_R}$ and $\psi_{e_R}$ would contain zero modes, while the other fields  would acquire a KK mass of the order of the compactification scale. Thus without the additional boundary masses that we 
need to add, the boundary conditions  would be 
\begin{equation} 
        \label{eq:flatBCsDN}
{\psi_{\nu_L}}_{|0,L}={\psi_{e_L}}_{|0,L}={\chi_{\nu_R}}_{|0,L}={\chi_{e_R}}_{|0,L}=0. 
\end{equation} 
We also assume that, as explained in \cite{CGMPT,CGPT}, on one brane the SU(2)$_L\times$ SU(2)$_R$ 
symmetry is broken to the diagonal SU(2)$_D$, while on the other brane SU(2)$_R\times$ U(1)$_{B-L}$ 
is broken to U(1)$_Y$. This means that on the SU(2)$_D$ brane the theory is non-chiral, and a 
Dirac mass term $M_D$ connecting the left and right fermions can be
added. 
Assuming that this Dirac mass term is added on the $y=L$ brane, 
the boundary conditions will be the same as in Appendix~\ref{sec:MD}:
\begin{eqnarray}
& \displaystyle
{\psi_{e_L}}_{|0^+}=0, \ \ \
{\chi_{e_R}}_{|0^+}=0,
\\
& \displaystyle
{\psi_{e_L}}_{|L^-} = -M_D L \, {\psi_{e_R}}_{|L^-},
\ \ \
{\chi_{e_R}}_{|L^-} = M_D L \, {\chi_{e_L}}_{|L^-}.
\end{eqnarray}
The KK decomposition is of the form~(\ref{eq:DiracKK}),
leading to an electron mass being solution of the equation:
\begin{equation} 
\tan (m_n L) =M_D L 
\end{equation} 
which, for $M_D\ll 1/L$, is solved by $m_0\sim M_D$. The lowest mass state is as expected a Dirac fermion with a mass that is just given by the Dirac mass added on the brane. 

However, the unbroken SU(2)$_R$ symmetry so far guarantees that the neutrino has the same mass
as the electron. The neutrino mass needs to be suppressed by some sort
of a see-saw mechanism, which can be achieved, as in Appendix~\ref{sec:BraneFermions}, 
by coupling the neutrino to a fermion localized on the brane where SU(2)$_R\times$ U(1)$_{B-L}$ 
is broken to U(1)$_Y$. Let us thus introduce an extra right-handed neutrino 
$\xi_{\nu_R}$ localized on that brane. Being SU(2)$_L \times$U(1)$_Y$ neutral, this
extra brane fermion can have a Majorana mass as well as a mixing mass term
with $\psi_{\nu_R}$ via the 4D Lagrangian at $y=0$
\begin{equation}
S_{4D}
=  
\int d^4 x  
\left(\vphantom{\frac{1}{2}}
- i \bar{\xi}_{\nu_R} \bar{\sigma}^\mu \partial_\mu \xi_{\nu_R}  
+ ML^{1/2} \left(\xi_{\nu_R} \psi_{\nu_R} + \bar{\xi}_{\nu_R} \bar{\psi}_{\nu_R} \right)
+ f \left( \xi_{\nu_R} \xi_{\nu_R}  + \bar{\xi}_{\nu_R} \bar{\xi}_{\nu_R} \right)
\right).
\end{equation}
The boundary conditions on the SU(2)$_L \times$ U(1)$_Y$ brane are then
\begin{eqnarray}
& \displaystyle
{\psi_{\nu_L}}_{|0^+}=0\, ,
\\
& \displaystyle
{\chi_{\nu_R}}_{|0^+} = - M L^{1/2} \xi_{\nu_R} \, ,
\\
& \displaystyle
-i \bar{\sigma}^\mu \partial_\mu {\chi_{\nu_R}}_{|0^+}
+ f  \, {\bar{\chi}_{\nu_R}}\vphantom{\chi_{\nu_R}}_{|0^+} 
- M^2 L \, {\bar{\psi}_{\nu_R}}\vphantom{\chi_{\nu_R}}_{|0^+} = 0\, .
\end{eqnarray}
The boundary conditions on the SU(2)$_D$ brane remain untouched:
\begin{eqnarray}
& \displaystyle
{\psi_{\nu_L}}_{|L^-} = -M_D L \, {\psi_{\nu_R}}_{|L^-} \, ,
\\
& \displaystyle
{\chi_{\nu_R}}_{|L^-} = M_D L \, {\chi_{\nu_L}}_{|L^-} \, .
\end{eqnarray}
The KK expansion will be of the form
\begin{eqnarray}
&\displaystyle
\xi = \sum_n c_n\, \xi_{n} (x),
\\
& \displaystyle
\chi_{\nu_L} = \sum_n g^{(\nu_L)}_n (z) \,  \xi_{n} (x), \ \
\chi_{\nu_R} = \sum_n g^{(\nu_R)}_n (z) \, \xi_{n} (x),
\\
& \displaystyle
\bar{\psi}_{\nu_L} = \sum_n f^{(\nu_L)}_n (z)\,  \bar{\xi}_{n} (x), \ \
\bar{\psi}_{\nu_R} = \sum_n f^{(\nu_R)}_n (z)\,  \bar{\xi}_{n} (x),
\end{eqnarray}
where the  $\xi_{n}$'s are 4D Majorana spinors of mass $m_n$:
\begin{equation} 
-i \bar{\sigma}^{\mu} \partial_\mu \xi_{n} + m_n\, \bar{\xi}_n = 0
\ \ \mathrm{and}\ \ 
-i \sigma^{\mu} \partial_\mu \bar{\xi}_n + m_n^*\, \xi_{n} = 0. 
\end{equation} 
Together with the bulk equations of motion these BC's lead
to the following mass spectrum for neutrino's (assuming the eigenmass $m_n$ real)
\begin{equation}
2\, \frac{f-m_n}{M^2 L} \left( M_D^2L \cos^2 m_nL - \sin^2 m_n L \right) =
 (1+M_D^2L^2) \sin 2m_n L \,.
\end{equation}
For $ f M_D^2 L\ll M^2$ and $M_D L\ll 1$ we get that the lowest mode is  a Majorana fermion with a mass 
approximately given by 
\begin{equation}
m_0 \sim \frac{f M_D^2 }{M^2},
\end{equation}
which is of the typical see-saw type since the Dirac mass, $M_D$, which is of the same order as that of the  electron mass, is suppressed by the large masses of the right handed neutrinos 
localized on the brane. Thus a realistic spectrum is achievable in this simple toy model for the leptons.

\subsection{Quark sector}

In order to get a realistic mass spectrum for the quarks, one cannot simply add a single brane localized 
two-component fermion as for the neutrinos, since in that case we would induce 
an anomaly in the effective theory. Instead, we need to introduce a vector-like brane localized 
color triplet with the quantum numbers of the up-type right handed quark and its conjugate 
(or the down type for a mixing for the down quarks). So the fields that we are considering now are 
\begin{equation} 
        \label{quark}
\left( \begin{array}{c} 
\chi_{u_L} \\ \bar{\psi}_{u_L} \\ \chi_{d_L} \\ \bar{\psi}_{d_L} 
\end{array} \right), 
\ \ 
\left( \begin{array}{c} 
\chi_{u_R} \\ \bar{\psi}_{u_R} \\ \chi_{d_R} \\ \bar{\psi}_{d_R} 
\end{array} \right), 
\ \ 
\left( \begin{array}{c} 
\xi_{u_R} \\ \bar{\eta}_{u_R} 
\end{array}\right), 
\end{equation} 
where $(\chi_{u_L},\chi_{d_L})$ will be identified as the SM $SU(2)_L$ quark doublet
and ${\psi}_{u_R}$  and ${\psi}_{d_R}$ as the SM right handed quarks; 
$({\psi}_{u_L}, {\psi}_{d_L})$ and $\chi_{u_R}$ and $\chi_{d_R}$ are
their partners needed to form complete 5D spinors and they will get KK masses of order
of the compactification scale. Finally $(\xi_{u_R},\bar{\eta}_{u_R})$ is a localized 4D Dirac spinor
that will couple to ${\psi}_{u_R}$ on the $SU(2)_L \times U(1)_Y$ brane at $y=0$.
We will thus again assume that in the absence of the brane 
localized mass terms and mixings the fields 
$\chi_{u_L}, \chi_{d_L}, \psi_{u_R}$ and $\psi_{d_R}$ 
would have zero modes. 
The 4D brane localized terms at $y=0$ are: 
\begin{eqnarray} 
&&\displaystyle
S_{y=0} = 
 \int d^4 x 
 \left( \vphantom{\frac{1}{2}}
 - i \bar{\xi}_{u_R} \bar{\sigma}^{\mu} \partial_\mu \xi_{u_R} 
 -i \eta_{u_R} \sigma^{\mu} \pd_\mu \bar{\eta}_{u_R} 
 \right. 
 \nonumber\\
&&\displaystyle
 \hspace{2cm}
 \left. \vphantom{\frac{1}{2}}
+ f \left( \eta_{u_R} \xi_{u_R} + \bar{\eta}_{u_R} \bar{\xi}_{u_R} \right)
+ M L^{1/2} \left( \xi_{u_R} \psi_{u_R} + \bar{\xi}_{u_R} \bar{\psi}_{u_R} \right)
\right) ,
\end{eqnarray} 
while at $y=L$, we just had an SU(2)$_D$ invariant Dirac mass term
mixing the left and the right quarks:
\begin{equation}
S_{y=L} = \int d^4 x\,M_DL  
\left(  \vphantom{\frac{1}{2}}
(\chi_{u_L} \psi_{u_R} + \chi_{u_R}\psi_{u_L} + 
 h.c.)
+
(\chi_{d_L} \psi_{d_R} + \chi_{d_R}\psi_{d_L} + 
 h.c. )
\right)
 \end{equation}
Since we have not included any mixing term at $y=0$ for the down-type quarks 
their spectrum will just be 
of the same form as for the electrons above, determined by the equation 
\begin{equation} 
\tan (m_n L) =M_D L. 
\end{equation} 
The boundary conditions for the up-type quarks 
are similar to those obtained in Appendix~\ref{sec:BraneFermions}:
\begin{eqnarray}
&\displaystyle
{\psi_{u_L}}_{|0^+} = 0,
\\
&\displaystyle
{\chi_{u_R}}_{|0^+} = - M L^{1/2}\, \xi_{u_R},
\\
&\displaystyle
(\partial^\mu \partial_\mu + f^2) {\chi_{u_R}}_{|0^+} 
-  M^2L\, i \sigma^\mu \partial_\mu \bar{\psi}_{u_R}{\vphantom{\psi_{u_L}}}_{|0^+} = 0,
\\
&\displaystyle
{\psi_{u_L}}_{|L^-} = -M_DL\, {\psi_{u_R}}_{|L^-},
\ \ \
{\chi_{u_R}}_{|L^-} = M_DL\, {\chi_{u_L}}_{|L^-}.
\end{eqnarray}
The KK decomposition will be of the usual form~(\ref{eq:DiracKK})
and leads to the following quantization equation similar to the neutrino's mass equation:
\begin{equation}   
2\, \frac{m_n^2-f^2}{M^2L m_n } 
\left(
 \sin^2 m_n L - M_D^2 L^2 \cos^2 m_n L 
 \right) 
 = 
 \left( 1 + M_D^2 L^2 \right)  \sin 2 m_n L,
\end{equation}
which again, for $ f M_D L \ll \sqrt{f^2+M^2}$ and $M_D L \ll 1$, has a  solution approximated by
\begin{equation} 
m_0 \sim \frac{M_D f}{\sqrt{f^2 + M^2}}.
\end{equation}  
If $M \ll f$, we can suppress the brane Dirac mass using a see-saw type 
mechanism, or if $M \sim f$ we can get a mass just slightly 
modified compared to the brane Dirac mass. We can use this freedom to generate
both the masses of the light first two generations as well as the masses of the massive third  generation.
We will discuss this much more in the context of the more realistic warped model in the coming section.

\section{Fermions in Warped Space} 
\setcounter{equation}{0} 

\subsection{Bulk equations of motion}

We now extend the discussion of the previous section to a truncated warped spacetime, 
which as usual we take to be a slice of AdS$_5$~\cite{RS}.  
The conformally flat metric corresponding to this situation is given by 
 \begin{equation} 
ds^2 = \left( \frac{R}{z} \right)^2 \left( \eta_{\mu \nu} dx^\mu dx^\nu - dz^2 \right). 
\end{equation} 
The boundaries of the spacetime are at $R\sim 1/M_{Pl}$ and $R^\prime\sim 1\,\mathrm{TeV}^{-1}$. Fermions in such a space have been considered  in~\cite{matthias,GherPom,HS1,nomurasmith,HSrecent}. Here we first briefly review the generic features of the fermion wave functions  in this space, and then repeat our analysis for 
the acceptable boundary conditions for this situation. 

The fermion action in a curved background is generically given by
\begin{equation}
        \label{eq:curvedaction}
S = \int d^5 x 
\sqrt{g}
\left( 
\frac{i}{2} (\bar{\Psi}\, e_a^M \gamma^a D_M \Psi 
- D_M \bar{\Psi} \, e_a^M \gamma^a \Psi)
-m \bar{\Psi} \Psi
\right),
\end{equation}
where $e_a^M$ is the generalization of the vierbein to higher dimensions (``f\"unfbein'') satisfying
\begin{equation}
e_a^M \eta^{ab} e^N_b = g^{MN},
\end{equation}
the $\gamma^a$'s are the usual Dirac matrices, and $D_M$ is the covariant derivative including the 
spin connection term.  Note again that the differential operators
have not been integrated by parts in order to avoid the introduction of any boundary terms that would otherwise be needed to make the action real.

For the AdS$_5$ metric in the conformal coordinates written above, 
$e_M^a =(R/z) \delta_M^a$, and
$D_\mu \Psi =( \partial_\mu+\gamma_\mu\gamma_5/(4z))\Psi$, $D_5 \Psi=\partial_5 \Psi$, 
however the spin connection terms involved in the two covariant derivatives of (\ref{eq:curvedaction})
cancel each other and thus do not contribute in total to the action. 
Finally, in terms of two component spinors, the action is  given by
\begin{equation} 
S = \int d^5 x 
\left(\frac{R}{z}\right)^4 
 \left( 
- i \bar{\chi}  \bar{\sigma}^\mu \partial_\mu \chi 
- i \psi  \sigma^{\mu} \partial_\mu \bar{\psi} 
+ \sfrac{1}{2} ( \psi \overleftrightarrow{\partial_5} \chi 
-  \bar{\chi}  \overleftrightarrow{\pd_5} \bar{\psi} )
+ \frac{c}{z} \left( \psi \chi + \bar{\chi} \bar{\psi} \right) 
\right), 
\end{equation} 
where $c$ is the bulk Dirac mass in units of the AdS curvature $1/R$, and again $\overleftrightarrow{\partial_5}  = \overrightarrow{\partial_5}-\overleftarrow{\partial_5}$
with the convention that the differential operators act only on the spinors and not on the metric factors.

The bulk equations of motion derived from this action are 
\begin{eqnarray} 
\label{bulkeq1}
-i \bar{\sigma}^{\mu} \pd_\mu \chi - \pd_5 \bar{\psi} + \frac{c+2}{z} \bar{\psi} = 0,
 \\ 
-i \sigma^{\mu} \pd_\mu \bar{\psi} + \pd_5 \chi + \frac{c-2}{z} \chi = 0.
\label{bulkeq2}
 \end{eqnarray} 
The KK decomposition takes its usual form (the case of a Majorana KK decomposition will be discussed in details in Section~\ref{sec:WarpedBraneMajoranaMass})
\begin{equation} 
\chi 
= 
\sum_n g_n (z) \chi_n (x)
\ \ \mathrm{and} \ \ 
\bar{\psi} 
=
\sum_n f_n (z) \bar{\psi}_{n} (x), 
\end{equation} 
where the 4D spinors $\chi_n $ and $\bar{\psi}_{n}$ satisfy the usual 4D Dirac equation with mass $m_n$:
\begin{equation} 
-i \bar{\sigma}^\mu \partial_\mu \chi_n + m_n \bar{\psi}_n = 0
\ \ \mathrm{and} \ \ 
-i \sigma^\mu \partial_\mu \bar{\psi}_n + m_n \chi_n = 0. 
\end{equation} 
The bulk equations then become ordinary (coupled) differential equations of first order
for the wavefunctions $f_n$ and $g_n$:
\begin{eqnarray} 
        \label{eq:beom1}
& \displaystyle
f^\prime_n + m_n g_n - \frac{c+2}{z} f_n = 0, 
\\
        \label{eq:beom2}
& \displaystyle
g^\prime_n  - m_n g_n + \frac{c-2}{z} g_n = 0. 
\end{eqnarray}  

For a zero mode, if the boundary conditions were to allow its presence, these bulk equations are already
decoupled and are thus easy to solve, leading to:
\begin{eqnarray} 
& \displaystyle
f_0  = C_0 \left( \frac{z}{R} \right)^{c+2}, 
\\
& \displaystyle
g_0 = A_0  \left( \frac{z}{R} \right)^{2-c},
\end{eqnarray}  
where $A_0$ and $C_0$ are two normalization constants
of mass dimension $1/2$.

For the massive modes, the first order differential equations can be uncoupled
and we obtain the two second order differential equations:
\begin{eqnarray} 
& \displaystyle
f^{\prime\prime}_n - \sfrac{4}{z} f^\prime_n + (m_n^2 - \sfrac{c^2-c-6}{z^2}) f_n = 0,
\\
& \displaystyle
g^{\prime\prime}_n - \sfrac{4}{z} g^\prime_n + (m_n^2 - \sfrac{c^2+c-6}{z^2}) g_n = 0,
\end{eqnarray}  
whose solutions are linear combinations of Bessel functions:
\begin{eqnarray} 
        \label{eq:Bessel1}
& \displaystyle
g_n(z) 
= 
z^\frac{5}{2} 
\left( 
A_n J_{c+\frac{1}{2}}(m_n z) + B_n Y_{c+\frac{1}{2}}(m_n z) 
\right)
\\ 
        \label{eq:Bessel2}
& \displaystyle
f_n(z) 
= 
z^\frac{5}{2} 
\left( C_n J_{c-\frac{1}{2}}(m_n z) + D_n Y_{c-\frac{1}{2}}(m_n z)
\right).
\end{eqnarray} 
The bulk equations of motion (\ref{eq:beom1})-(\ref{eq:beom2}) further impose that
\begin{equation}
A_n = C_n 
\ \ \mathrm{and} \ \ 
B_n = D_n.
\end{equation}
%

\subsection{Boundary conditions}

To find the consistent boundary conditions, we need to again consider the boundary terms in the variation of the action: 
\begin{equation} 
\delta S_{bound} = 
\frac{1}{2}
\int d^4x 
\left[
\frac{R^4}{z^4}
(\delta \chi \, \psi 
- \delta\psi \, \chi
-\delta\bar{\psi} \, \bar{\chi} 
+\delta \bar{\chi} \, \bar{\psi})
\right]_R^{R'}, 
\end{equation} 
which agrees with the expression for flat space up to the irrelevant factor of $R^4/z^4$. 
Thus the boundary conditions that make the boundary variation of the action vanish will 
generically be of the  same form as for the flat space case, 
that is those given in equations (\ref{bc1})-(\ref{bc2}). If the fermions 
are in real representations of the gauge group $M_\alpha^{\,\beta}$ 
is allowed to be non-vanishing, but if  they are in complex representations $M_\alpha^{\,\beta}$ has to be zero. 

\subsubsection{Boundary conditions in absence of extra boundary operators}

As an example let us consider the simplest case, when we make the conventional choice of
imposing Dirichlet BC's on both ends: 
\begin{equation} 
\psi_{|R^+} = 0
\ \ \mathrm{and} \ \
\psi_{|R^{\prime\, -}} = 0. 
\end{equation}  
These BC's allow for a chiral zero mode in the $\chi$ sector while the profile
for $\psi$ has to be vanishing, so we find for an arbitrary value of the bulk mass $c$ that the zero modes
are given by:
\begin{equation} 
f_0 = 0
\ \ \mathrm{and} \ \
g_0 = A_0 \left(  \frac{z}{R} \right)^{2-c}. 
\end{equation}  
The main impact $c$ has on the zero mode is where it is localized, 
close to the Planck brane (around $z=R$) or the TeV brane (around $z=R'$). This can be seen 
by considering the normalization of the fermion wave functions.
To obtain a canonically normalized 4D kinetic term for the zero mode, one needs
\begin{equation}
        \label{eq:normM0}
\int_R^{R'} dz 
\left(\frac{R}{z} \right)^5 \frac{z}{R} \, A_0^2  \left(  \frac{z}{R}  \right)^{4-2c} 
=1
\ \ i.e. \ \
A_0= \frac{\sqrt{1-2c}}{R^c\sqrt{R^{\prime\, 1-2c}-R^{1-2c}}},
\end{equation}
where the first factor in the integral comes from the volume $\sqrt{g}$, the $z/R$ factor from the vierbein and the rest is the wave function itself (squared). To figure out where this zero mode is localized in a covariant
way, we can send either brane to infinity and see whether the zero mode remains
normalizable. For instance, sending the TeV brane to infinity, $R'\to \infty$, 
the integral (\ref{eq:normM0}) converges only for $c>1/2$, in which case the zero mode
is localized near the Planck brane. Conversely, for $c<1/2$, when the Planck brane is sent
to infinity, $R \to 0$, the integral (\ref{eq:normM0}) remains convergent
and the zero mode is thus localized near the TeV brane.
In the AdS/CFT language~\cite{holography} this corresponds to the fact that for $c>1/2$ the fermions will be elementary
(since they are localized on the Planck brane), while for $c<1/2$ they are to be considered as composite
bound states  of the  CFT modes (since they are peaked on the TeV brane). 

This result can also be seen easily when using the proper distance coordinate along the extra dimension. In this case the AdS$_5$ metric is written as  ($k=1/R$ is the AdS curvature):
\begin{equation}
ds^2 = e^{-2 k y} \, \eta_{\mu\nu} dx^\mu dx^\nu - dy^2.
\end{equation}
And the actual normalized wavefunctions
(including the volume and vierbein factors) are 
\begin{eqnarray}
e^{-(2c-1)k(y-y_{Pl})}  & {\rm for} \ c>1/2 \ ({\rm and} \ y_{TeV} \to \infty),
\\
e^{-(1-2c)k(y_{TeV}-y)}  & {\rm for} \ c<1/2 \ ({\rm and} \ y_{Pl} \to -\infty).
\end{eqnarray}

Finally, let us point out that if we were to impose Dirichlet BC's on both ends of the interval
for $\chi$, we would have found a zero mode in the $\psi$ sector. And this zero mode would have been
localized on the Planck brane for $c<-1/2$ and localized
on the TeV brane for $c>-1/2$.

\subsubsection{Boundary conditions with a brane Majorana mass term}
\label{sec:WarpedBraneMajoranaMass}

To familiarize ourselves more with the BC's in warped space, 
we will repeat the flat-case analysis of section~\ref{sec:FlatBraneMajoranaMass}
and  consider the case when a Majorana mass is added on the Planck brane for the $\chi$ field 
(which would otherwise have a zero mode).\footnote{While this paper was 
in preparation \cite{HSrecent} appeared, which also presents 
a detailed treatment of a Planck-brane localized Majorana mass term.} 
Based on our discussions we expect that the boundary condition
on the Planck brane would be modified to 
\begin{equation}
        \label{majoranabc}
\left( \cos \alpha \, \psi -\sin \alpha \, \chi\right)_{|R^+} 
= 0
\end{equation}
where $\sin \alpha =0$ corresponds to the case with no Majorana mass, while $\cos\alpha =0$ to the case with a very large Majorana mass. 
To identify the actual relation between the Majorana mass and $\alpha$ 
we again consider adding the Majorana mass at $z=R+\epsilon$, read off the BC's from the bulk 
equations, and then send $\epsilon\to 0^+$. In this case the bulk equation of motion will be modified to
\begin{equation}
-i \bar{\sigma}^{\mu} \partial_\mu \chi 
- \partial_5 \bar{\psi} 
+ \frac{c+2}{z} \bar{\psi} 
+ \frac{M^*R^2}{z}\bar{\chi} \, \delta (z-R-\epsilon ) 
= 0,
\end{equation}
where $M$ is the Majorana mass added. There will be a discontinuity
in the profile for $\psi$, it is given by the jump equation
\begin{equation}
[\psi]_{|R+\epsilon}= MR \, \chi_{|R+\epsilon},
\end{equation}
from which, using $\psi_{|R}=0$, we can read off the relevant boundary condition
\begin{equation}
\psi_{|R^+} = MR\, \chi_{|R^+},
\end{equation}
which is indeed of the form (\ref{majoranabc}). 

The KK decomposition is of the form
\begin{equation} 
\chi 
= 
\sum_n g_n (z) \xi_n (x)
\ \ \mathrm{and} \ \ 
\bar{\psi} 
=
\sum_n f_n (z) \bar{\xi}_{n} (x), 
\end{equation} 
where the 4D spinors $\xi_n $  satisfy the usual 4D Majorana equation with mass $m_n$:
\begin{equation} 
-i \bar{\sigma}^\mu \partial_\mu \xi_n + m_n \bar{\xi}_n = 0
\ \ \mathrm{and} \ \ 
-i \sigma^\mu \partial_\mu \bar{\xi}_n + m_n^* \xi_n = 0. 
\end{equation} 
Instead of the expansion in the Bessel and Neumann functions, it turns out that it is more
convenient to expand in terms of $J_\nu$ and $J_{-\nu}$. These functions are linearly 
independent as long as $\nu$ is not an integer, that is if $c\neq 1/2 +$ integer. We will
then treat the $c=1/2$ as a special case later. The reason why it is more convenient to 
use these functions is that the expansion for small arguments, needed for an approximate 
solution for the lowest modes, will be much simpler if we use this basis.
Thus the wave functions $f_n$ and $g_n$ will be of the form 
\begin{eqnarray} 
& \displaystyle
g_n(z) 
= 
z^\frac{5}{2} 
\left( 
A_n J_{c+\frac{1}{2}}( |m_n| z) + B_n J_{-c-\frac{1}{2}}( |m_n| z) 
\right),
\\ 
& \displaystyle
f_n(z) 
= 
z^\frac{5}{2} 
\left( C_n J_{c-\frac{1}{2}}( |m_n| z) + D_n J_{-c+\frac{1}{2}}( |m_n| z)
\right).
\end{eqnarray} 
and the bulk equations further require that
\begin{equation}
m_n A_n = |m_n| C_n
\ \ \mathrm{and} \ \
m_n B_n = -|m_n| D_n. 
\end{equation}
The two boundary conditions
\begin{equation}
\psi_{|R^+}= MR \, \chi_{|R^+} 
\ \ \mathrm{and} \ \
\psi_{|R^{\prime\,-}}=0
\end{equation}
then lead to the equation determining the eigenvalues $m_n$:
\begin{equation}
J_{c-\frac{1}{2}} \tilde{J}_{-c+\frac{1}{2}}
- J_{-c+\frac{1}{2}}  \tilde{J}_{c-\frac{1}{2}} 
=
\pm MR \, (
J_{c+\frac{1}{2}}  \tilde{J}_{-c+\frac{1}{2}} 
+ J_{-c-\frac{1}{2}}  \tilde{J}_{c-\frac{1}{2}} 
),
\end{equation}
where $J_\nu=J_\nu (|m_n|R)$ and $\tilde{J}_\nu=J_\nu (|m_n|R')$.
This equation can be approximately solved for the lowest eigenmode (assuming that $m_0 R' \ll 1$)
by expanding the Bessel functions for small arguments as
\begin{equation}
J_\nu (x)\sim \left(\frac{x}{2}\right)^\nu \frac{1}{\Gamma (\nu +1)}.
\end{equation}
We find, that for $c>1/2$ the lowest eigenmode is approximately given by
\begin{equation}
m_0 \sim  (2c-1)M,
\end{equation}
while for $c<1/2$ it is 
\begin{equation}
m_0 \sim (1-2c)M \left(\frac{R}{R'}\right)^{1-2c}.
\end{equation}
The $c=1/2$ case has to be treated separately, since in that case the expansion has to be in terms
of the Bessel and Neumann functions. The equation that the eigenvalues have to solve will 
be given by
\begin{equation}
J_0 \tilde{Y}_0 - Y_0 \tilde{J}_0 
= \pm MR \, (
J_1 \tilde{Y}_0 - Y_1 \tilde{J}_0
).
\end{equation}
For the lightest mode we find
\begin{equation}
m_0 \sim \frac{M}{\log{\frac{R'}{R}}}.
\end{equation}
The interpretation of these expressions is quite clear. When $c>1/2$, we expect the resulting mass 
to be proportional to the mass added on the Planck brane, since the fields themselves are localized 
near the Planck brane. For $c<1/2$ the zero mode is localized near the TeV brane, so adding a mass on 
the Planck brane has only a small effect due to the wave function suppression. For the $c=1/2$ case
the wave function is flat, and one expects 
the usual volume suppression as in flat backgrounds; that is one
expects a suppression by the proper distance between the branes.
The expressions above are in 
clear correspondence with these expectations.

\section{Fermion Masses in the Higgsless Model of Electroweak Symmetry Breaking in Warped Space} 
\label{sec:WarpedNoHiggs}
\setcounter{equation}{0} 

We are now finally ready to consider the SU(2)$_L\times$SU(2)$_R\times$U(1)$_{B-L}$ model
in warped space, where electroweak symmetry breaking is achieved by boundary conditions (rather
than by a Higgs on the TeV brane). As discussed in~\cite{CGPT}, this model has a custodial 
SU(2) symmetry that protects the $\rho$ parameter from large corrections, and thus to leading log 
order  the structure of the standard model in the gauge sector is reproduced. However, an obvious
lingering question  is whether realistic values of the fermion mass could be obtained 
in this model. This model could be considered the AdS dual of walking technicolor \cite{walking}, and it is well-known
that in technicolor theories it is difficult to naturally obtain a realistic fermion spectrum. Here we will
show, that in the extra dimensional model there is enough freedom in the parameter space of the 
theory to be able to incorporate the observed fermion masses. 

\subsection{Quark sector}

The left handed leptons and quarks will be in SU(2)$_L$ doublets, while the right handed ones in
SU(2)$_R$ doublets, exactly as in (\ref{leftlep}),(\ref{rightlep}) and (\ref{quark}).
Thus we will have two SU(2) doublet Dirac fermions for the leptons and 
two separately for the quarks in the bulk for every generation, 
$\chi_{L,R}, \bar \psi_{L,R}$. 
Each Dirac fermion has a bulk mass $c_{L,R}$ and a Dirac mass $M_D$ 
on the TeV brane that mixes the two bulk fermions.  In addition we assume that there is 
a Dirac fermion localized on the Planck brane that mixes with $\psi_R$. This will again be 
necessary to be able to sufficiently split the masses of the up and down type fermions. 
 
We again assume that the fields $\psi_{L,R}$ and $\chi_{L,R}$ are such that 
in the absence of brane localized masses/mixings the fields $\chi_L$ and $\psi_R$ would have
zero modes, that is the BC's in the absence of the brane terms are as in (\ref{eq:flatBCsDN})
\begin{equation}
{\psi_L}_{|R,R'}={\chi_R}_{|R,R'}=0.
\end{equation}
Since we would like the zero modes (at least for the light fermions) to be localized near the Planck
brane (in order to recover the SM relations for the gauge couplings), we need to pick
the bulk mass terms $c_L>1/2$ and $c_R<-1/2$. The reversal of the inequality for $c_R$ is due to the 
fact that for the right handed doublets we want the $\psi$ fields to have zero modes, and for these 
types of zero modes the localization properties as a function of $c$ are modified, with 
$c<-1/2$ localized near the Planck brane while $c>-1/2$ near the TeV brane.

The bulk part of the fermion action will be as in (\ref{eq:curvedaction}), and the bulk equations of motion are as
in (\ref{bulkeq1})-(\ref{bulkeq2}). To find the appropriate boundary conditions, we need to consider the brane 
localized mass and mixing terms. The mixing term on the Planck brane will be of the form 
\begin{equation}
        \label{Planckinduced}
S_{Pl}=
\int d^4 x 
\left( \vphantom{\frac{1}{2}}
 -i \bar{\xi}  \bar{\sigma}^\mu \partial_\mu \xi -i 
\eta \sigma^{\mu} \partial_\mu \bar{\eta} 
+f\left( \eta \xi + \bar{\xi} \bar{\eta} \right)  +M \sqrt{R}\left( \psi_{R} \xi+ \bar{\xi} \bar{\psi}_R \right)
\right)_{|z=R},
\end{equation}
where $\xi$ and $\eta$ are brane localized fermions, which together form a Dirac fermion with a Dirac mass $f$
on the brane. On the 
Planck brane only SU(2)$_L\times$U(1)$_Y$ is unbroken, and this extra Dirac fermion is assumed to be an SU(2)$_L$ 
singlet carrying the U(1)$_Y$ quantum numbers of the right handed SM fermion fields, such that the mixing 
with mixing mass $M$ in (\ref{Planckinduced}) is allowed.  This is the analog of the $y=0$ term in the 
Lagrangian in the flat space case discussed in Sections 4 and 5. Since the warp factor on the Planck brane is one,
the boundary conditions following from this brane localized Lagrangian will exactly match that in flat space:
\begin{eqnarray}
(\partial^\mu \partial_\mu+f^2) {\chi_R}_{|R^+}
&=&  M^2 R \, i \sigma^\mu\partial_\mu {\psi_R}_{|R^+}, 
        \label{planck} 
\\ 
{\psi_L}_{|R^+} &=& 0 .
 \end{eqnarray}
For modes with $m_n^2 \ll f^2$  the BC (\ref{planck}) can be approximated by 
\beq 
{\chi_R}_{|R^+}  =m_n  \frac{M^2 R}{f^2}  \, {\psi_R}_{|R^+}, 
\eeq 
where $m_n$ is the n$^{th}$ mass eigenvalue.

The mass term on the TeV brane will be given by
\begin{equation}
S_{TeV}= 
\int d^4 x\, 
\left(\frac{R}{z}\right)^4 M_D R' 
\left( \vphantom{\frac{1}{2}}
\psi_R \chi_{L} + \bar{\chi}_L \bar{\psi}_R+ 
\psi_L \chi_{R} + \bar{\chi}_R \bar{\psi}_L 
\right)_{|z=R^\prime} .
\end{equation}
Note, that in order to have a natural Lagrangian the parameter $M_D R'$ should be of order one,
thus $M_D$ should be of the order of TeV. From this mass term the boundary conditions on the TeV brane will be analogous to the flat space case discussed 
in Sections 4 and 5: 
\beq 
&& \displaystyle
{\psi_L}_{|R^{\prime\,-}}
=
-M_D R^\prime \, {\psi_R}_{|R^{\prime\, -}} 
        \label{TeV} \\ 
&& \displaystyle
{\chi_R}_{|R^{\prime\, -}} 
=  
M_D R^\prime \, {\chi_L}_{|R^{\prime\, -}} 
 \eeq 

For the mode decomposition the bulk wavefunction solutions we take for the general case
$1/2 + c_{L,R}\ne$ integer: 
\begin{eqnarray} 
\chi_{L,R} 
&=&  
z^\sfrac{5}{2}
\left( 
A^n_{L,R} J_{\sfrac{1}{2}+c_{L,R}}(m_n z)
+ B^n_{L,R} J_{-\sfrac{1}{2}-c_{L,R}}(m_n z) 
\right) ,
\nonumber \\ 
\psi_{L,R} 
&=& 
z^\sfrac{5}{2}
\left( 
A^n_{L,R}  J_{-\sfrac{1}{2}+c_{L,R}}(m_n z) 
-  B^n_{L,R} J_{\sfrac{1}{2}-c_{L,R}}(m_n z) 
\right),
\end{eqnarray} 
where $m_n$ is the 4D mass of the given mode that one is considering. 
For a mode with $m_n R^\prime \ll 1$ we can expand the Bessel functions for small arguments, and since the
coefficients $A_{L,R},B_{L,R}$ depend on the eigenvalue $m_n$, some overall powers of $m_n$
can be absorbed into these constants to make the expansion more transparent (from here on we will  
suppress the index $n$), keeping the terms at most quadratic in $m_n$, we get:
\begin{eqnarray} 
\chi_{L,R} 
&=&   
z^{2}
\left(
\frac{\tilde{A}_{L,R} m z^{c_{L,R}+1}}{2c_{L,R}+1} 
+ \tilde{B}_{L,R} z^{-c_{L,R}}\left(1-\frac{m^2z^2}{2-4c_{L,R}}\right) 
\right), 
\nonumber \\ 
\psi_{L,R} 
&=& 
z^{2}
\left(
\tilde{A}_{L,R}  z^{c_{L,R}}\left(1-\frac{m^2z^2}{2+4c_{L,R}}\right) 
-\frac{\tilde{B}_{L,R} m z^{1-c_{L,R}}}{1-2c_{L,R}} 
\right).
\end{eqnarray} 
Imposing the above boundary conditions we find that the lightest eigenmode is approximately
given by\footnote{In the following formulae we only keep the terms which
  can be leading in $R'/R$ for  chosen values of bulk masses. Among the
  remaining terms we separately keep only the leading contributions
  proportional to $M^2$ and $f^2$ terms because $M/f$ is a free parameter
  which may be large.} (assuming again 
$c_L>1/2, c_R<-1/2$)
\begin{equation}
\label{generalmass}
m_0 \sim
\frac{\sqrt{(2c_L-1)(-2c_R-1)}f M_D}{\sqrt{f^2+(-2c_R-1)M^2}} \left(\frac{R}{R'}\right)^{c_L-c_R-1}.
\end{equation}
The approximation $c_L>1/2$, $c_R<-1/2$ should be sufficient for the light quarks, but for the top quark we need 
the wavefunctions to have a larger overlap on the TeV brane. Therefore we need to consider the case
 $c_L<1/2$, $c_R>-1/2$ so that the top quark is localized near the TeV brane rather than near the Planck brane,
which will make it possible to get a sufficiently large top quark mass. Using similar methods as before we find
that in this case ( $c_L<1/2$, $c_R>-1/2$) the approximate lowest mass eigenvalue is
\begin{equation}
m_0^2\sim \frac{M_D^2 (1-2c_L)(1+2 c_R)}{1+M_D^2 R'^2 (1-c_L+c_R)+
\left(\frac{R}{R'}\right)^{2c_R+1}
\frac{M^2}{f^2}(1+2c_R)\left( 1+\frac{(1-2c_L)M_D^2 R'^2}{1-2c_R}\right)}~.
       \label{tevlocalized}
\end{equation}

 For completeness we also briefly discuss the special cases when $c_L=1/2$ and/or $c_R=1/2$. 
With $c_L=1/2$, $c_R>-1/2$ we find that the lightest eigenmode is 
given by 
\beq
\label{specialmass}
m_0^2
\sim
\frac{2(1+2c_R) f^2M_D^2 R'^{1+2 c_R}}{\log \frac{R'}{R}
\left(
    R'^{1+2c_R}f^2 (2+(1+2c_R)M_D^2 R'^2)+2R^{1+2c_R}((1+2c_R)M^2-f^2)
\right)}
~.
\eeq
The complementary case $c_R=-1/2$ and $c_L>1/2$ gives
\begin{equation}
m_0^2
\sim
\frac{(2c_L-1) f^2 M_D^2}{f^2 \log \frac{R^\prime}{R}+M^2} \left(\frac{R}{R'}\right)^{2c_L-1}~,
\end{equation}
while for the doubly special case $c_L=-c_R=1/2$: 
\begin{equation}
m_0^2
\sim
\frac{f^2 M_D^2}{\log\frac{R'}{R}
\left(
f^2\log\frac{R'}{R}+M^2
\right)}~.
\end{equation}

Let us now use these expressions to demonstrate that it is possible to obtain a 
realistic mass spectrum for all the SM fermions. 
We will use (\ref{generalmass}) for the first two generations, while for the third   
generation quarks we use (\ref{tevlocalized}). We have also numerically solved the bulk equations 
with the appropriate boundary conditions, and found that (\ref{generalmass})-(\ref{tevlocalized}) 
are generically good approximations for the lowest eigenvalues, up to the ten percent level. For the results 
presented below we have used the numerical solutions to the eigenvalue equations rather than the 
approximate formulae (\ref{generalmass})-(\ref{tevlocalized}).
We also use the numerical solution to find the 
lightest KK excitations in each case. 
We will not attempt to explain all the observed CKM matrix elements in this 
paper, though we see no reason why it should be hard to obtain the right values. In order
to correctly reproduce~\cite{CGPT}
the masses of the $W$ and the $Z$, throughout the 
fits we will use the values  $R=10^{-19}$~GeV$^{-1}$, $R^\prime=2\cdot 10^{-3}$~GeV$^{-1}$. The parameter $M_D$ 
should be of the order of the TeV scale, while the splitting between the up and down-type fermions 
will be obtained by choosing an appropriate value for the ratio $M/f$, that is the ratio of the 
mixing with the brane fermions to the diagonal mass of the brane fermions. A reasonable quark mass spectrum
can be obtained using the following parameters:

(i) For the first generation take   
$c_L=-c_R=.6$, $M_D=50$ GeV and $M/f=3.8$ for the up sector and $M/f=0$ for the down sector. Then $m_u\approx 3$ MeV, $m_d\approx 6$ MeV
and the first KK excitations appear at 1.2 TeV then 1.3 TeV (both for up and down). 

(ii) For the second generation take 
$c_L=-c_R=.52$, $M_D=112$ GeV and $M/f=50$ for the strange sector. Then $m_s\approx 110$ MeV, $m_c\approx 1.3$ GeV
and the first KK excitations appear at 1.1 TeV then 1.3 TeV (both for s and c). 

(iii) For the third generation we need localize both the left handed and the right handed zero modes
near the TeV brane in order to be able to get a large enough top quark mass. One numerical example is
$c_L=0.4, c_R=-1/3$, $M_D=900$ GeV, $f=2.5 \cdot 10^{10}$ GeV and $M=10^{15}$ GeV for the bottom sector and $M=0$ for the top sector. For these 
parameters we get $m_{top}\approx 175$~GeV and $m_{b}\approx~4.5$ GeV.  The first KK excitations of the bottom
quark appear at the relatively low value of $\sim 550$~GeV, while for the top quark at $\sim 700$~GeV.
This would imply that the third generation (since it would be localized near the TeV brane) 
would be very different from the first two, and interesting effects in flavor physics could be 
observable. For a recent analysis of examples of the consequences for a composite third 
generation see~\cite{Burdman}. 

\begin{figure}[h!t]
\centerline{\includegraphics[width=0.3\hsize]{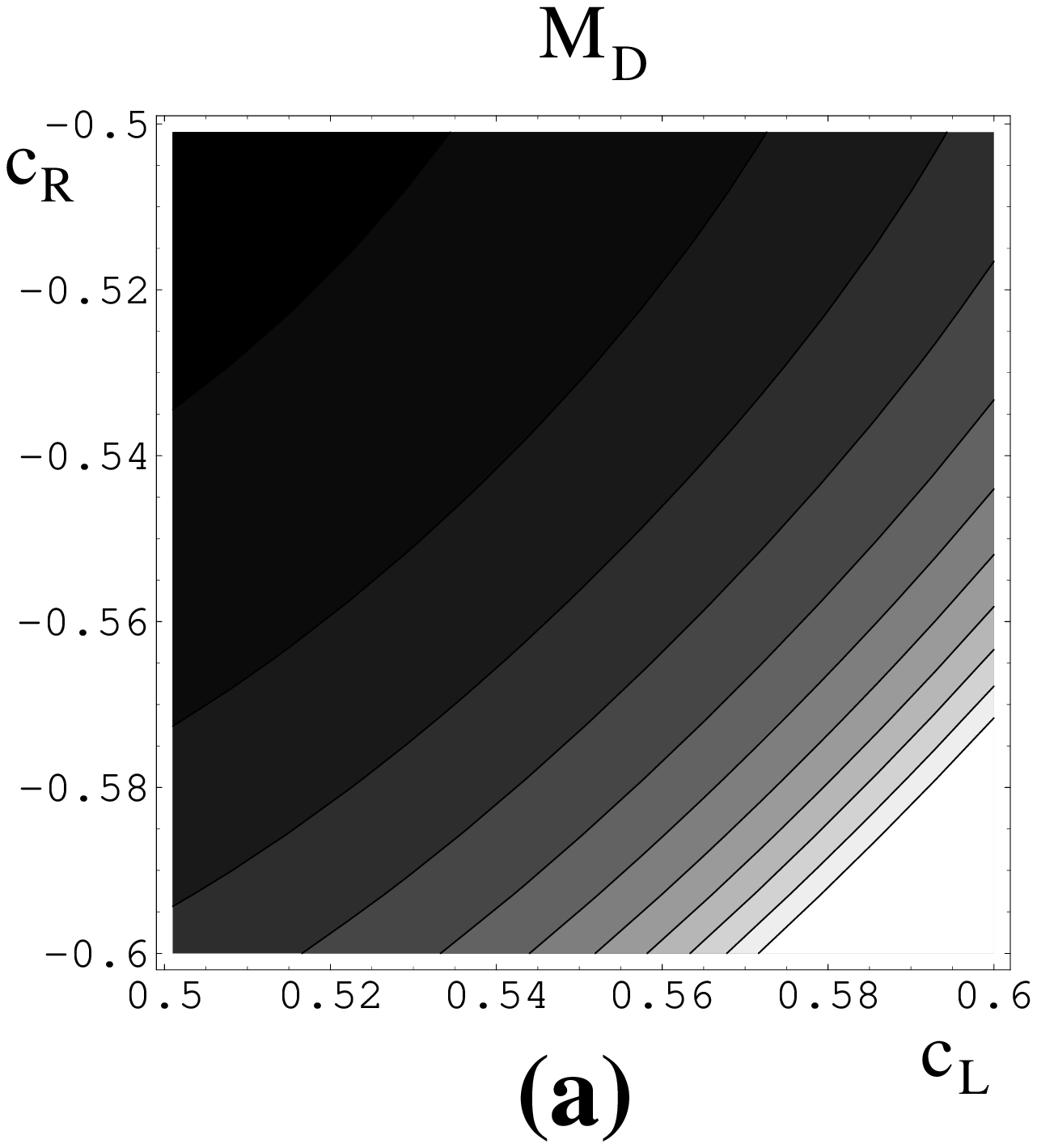}
\hspace{2cm}
\includegraphics[width=0.3\hsize]{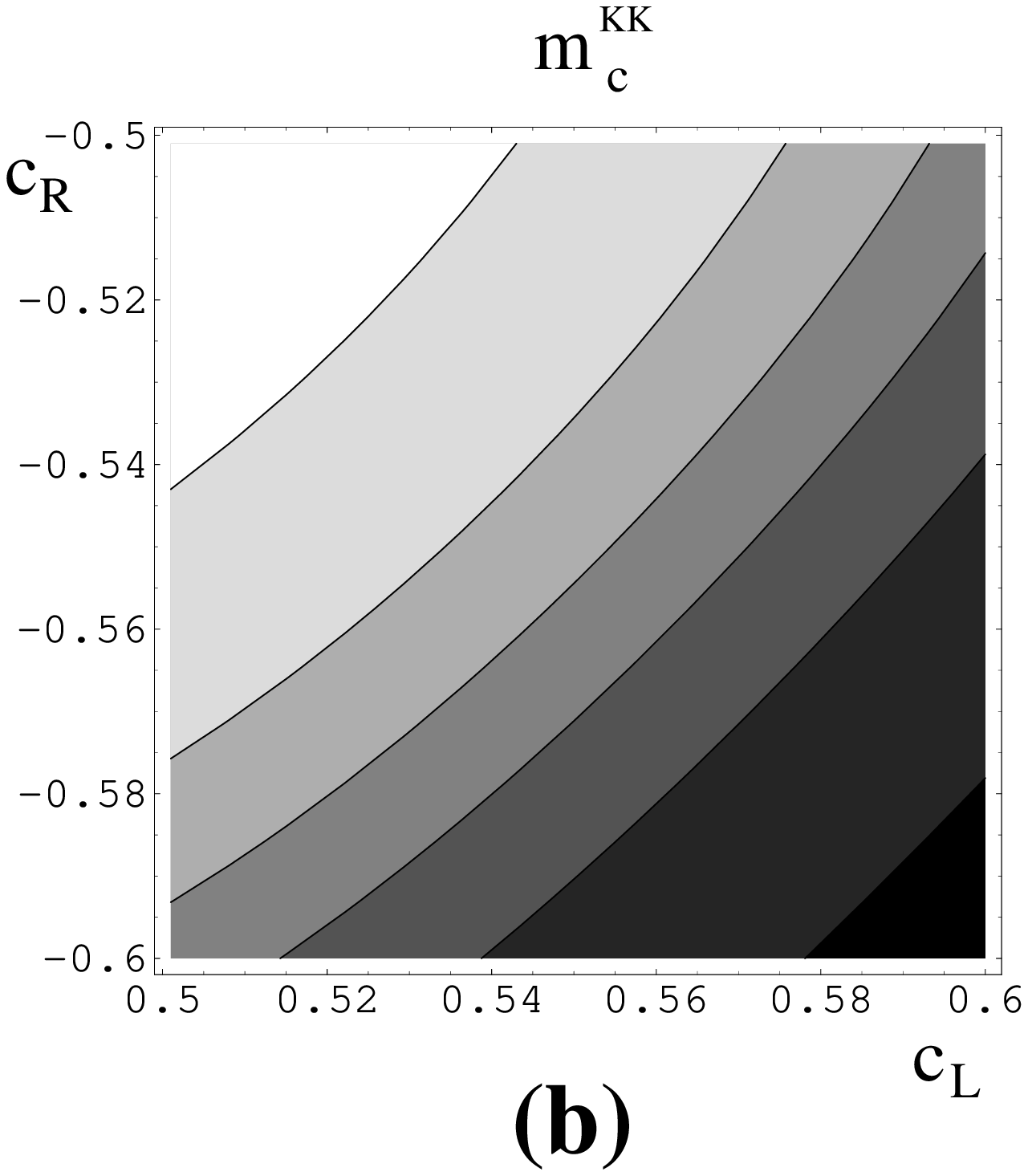}}
\caption{\label{fig:varyingc} {\bf (a)} Contour plot of the value of $M_D$ needed to obtain $m_{c}=1.2$ GeV forvarying values of $c_L$ and $c_R$. The regions starting with the darkest moving toward the 
lighter ones correspond to $M_D=0.1,0.3,0.6,1,1.5,2,2.5,3,3.5,4,4.5$ TeV.
{\bf (b)} Contour plot of the value of the lightest KK mass 
for the second generation quarks assuming that $M_D$ is chosen such that $m_{c}=1.2$ GeV, for
varying values of $c_L$ and $c_R$. The regions starting with the darkest moving toward the 
lighter ones correspond to $M_{KK}=0.1,0.3,0.5,0.7,0.9,1.1$ TeV
}
\end{figure}

The above numbers are only given for the purpose of demonstrating the viability of obtaining a realistic set 
of quark masses. However, there are several free parameters that one can vary for 
obtaining the correct masses: $c_L,c_R$ and  $M_D$, while the ratio $M/f$ is mostly set by the amount of 
splitting within a multiplet. Here we have only assumed the simplest possibility when one of the two 
fermions within a generation have a mixing with the brane localized fermions. Clearly, there is a much 
richer spectrum of possibilities for such mixings which we will however not deal with here. To illustrate 
some of the available free parameters, we have varied the $c$'s and $M_D$ around the solution
for the second generation, while keeping the c and s quark masses fixed. The resulting relation between
$c_L,c_R$ and $M_D$ is displayed in Fig.~\ref{fig:varyingc}(a). In Fig.~\ref{fig:varyingc}(b) 
we show the dependence of the 
mass of the lightest  KK mode on the parameters $c_L$ and $c_R$. 
Note, that a characteristic feature of all of these 
solutions is that the mass of the lightest KK mode decreases with increasing $M_D$. This is due to the fact
that  for large $M_D$ the KK mass is sensitive to the value of the $c$'s but not $M_D$ itself. This can
be simply seen by taking the large $M_D$ limit, where the resonance mass is just set by the
scale $1/R^\prime$.

\begin{figure}
\centerline{\includegraphics[width=0.3\hsize]{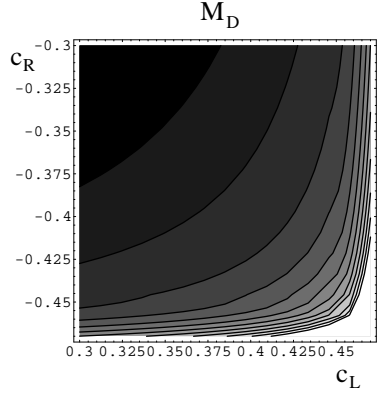}}
\caption{\label{fig:topmass} Contour plot of the value of $M_D$ needed to obtain $m_{top}=175$ GeV for
varying values of $c_L$ and $c_R$. The regions starting with the darkest moving toward the 
lighter ones correspond to $M_D=1,1.5,2,\ldots ,6,6.5$ TeV.}
\end{figure}

Since the amount of interesting new flavor physics in the third generation crucially depends 
on the deviation of $c_L$ from 1/2, we have examined how small $c_L-1/2$ could be. For this we have generated
calculated the acceptable values of $c_L,c_R$ and $M_D$ that would give us the correct top quark mass,
which is shown in Fig.~\ref{fig:topmass}. One can see, that the smallest possible value for $1/2-c_L$ is 
$\sim$0.03.  

It is also interesting to note that since the mixing with very heavy fermions on the Planck brane is
essentially equivalent to introducing brane kinetic terms, we can easily implement the 
Hiller--Schmaltz mechanism for solving the strong CP problem \cite{Hiller}.  If all intergeneration mixing
arises on the Planck brane by mixing with the heavy Planck brane fermions, then the net effect for the 
light fermions is that all the mixing appears in kinetic terms, then all complex phases  can be rotated
into the CKM matrix without introducing strong CP violation \cite{Hiller}.

\subsection{Lepton sector}

One has a variety of options for generating the lepton masses. The nicest 
possibility would be to have an extra dimensional 
implementation of the usual neutrino see-saw mechanism which takes advantage of the fact that the 
right handed neutrino is a singlet under all the SM gauge groups. This implies that on the Planck 
brane one can simply add a brane localized mass term to the right handed neutrino (since SU(2)$_R$ is 
broken there). Using similar methods as before we find that the neutrino mass is given by 
\begin{equation}
        \label{neutrinoseesaw}
m_\nu=(2c_L-1) \frac{M_D^2}{M_R} \left(\frac{R}{R'}\right)^{2(c_L-c_R-1)},
\end{equation}
where $M_R$ is the Majorana mass of the right handed neutrino on the Planck brane. In absence of any brane localized fermions,  the charged lepton 
masses are given by (\ref{generalmass}) with $M=0$ then we find the relation between the neutrino and
the charged lepton mass to be
\begin{equation}
m_{\nu_i}= \frac{1}{-2c_R-1}\frac{m_{l^-_i}^2}{M_R},
\end{equation}
where the $m_{l^-_i}$ are the masses of the charged leptons. Note, that this is almost completely 
analogous to the usual see-saw formula, with the only difference being that the large scale directly 
suppresses the charged lepton mass squares, and not a mass of order 100 GeV as is usually assumed. That is, the difference
is in the appearance of the charged lepton Yukawa coupling. Using this mechanism one could
get realistic lepton masses using for example the following parameters:

(i) For the first generation take   
$c_L=-c_R=.65$, $M_D=130$ GeV and $M_R=10^{10}$~GeV. Then $m_e \approx 500$ keV, $m_{\nu_e}\approx  10^{-7}$ eV.

(ii) For the second generation take 
$c_L=-c_R=.58$, $M_D=250$ GeV and $M_R=3\cdot 10^{10}$~GeV. Then $m_\mu\approx 100$ MeV, $m_{\nu_\mu}\approx 2 
\cdot 10^{-3}$ eV.

(iii) For the third generation take 
$c_L=-c_R=.53$, $M_D=240$ GeV and $M_R= 10^{12}$~GeV. Then $m_\tau \approx 1.7$ GeV, $m_{\nu_\tau}\approx 4 \cdot
10^{-2}$ eV.

Instead of the appearance of the intermediate scale $M_R \sim 10^{11}$ GeV for the Majorana mass of the 
right-handed neutrino (which is somewhat lower than usually assumed), one can insist on the real 
see-saw formula. A real see-saw mechanism can actually be achieved if
the suppression of the charged lepton masses compared to $M_D$ is 
the  consequence not only of a warp factor suppression but also of an additional
mixing with Planck-brane localized fermions as in the general case (\ref{generalmass}). The 
neutrino masses would still be given by (\ref{neutrinoseesaw}), with a scale $M_R$ different from
the scale $M$ in (\ref{generalmass}). This way one could choose an $M_D$ of order .1 to 2~TeV for all
three generations, and the scale $M_R$ needed for the neutrino masses would be closer to the 
usually assumed values of order $10^{15}$ to $10^{16}$~GeV. A possible choice of parameters could be 
for example:

(i) For the first generation take   
$c_L=-c_R=.55$, $M_D=100$ GeV, $M/f=1500$ for the suppression of the electron mass to $m_e\approx 500$ keV,
and $M_R = 10^{16}$ GeV for the suppression of the electron neutrino mass to $m_{\nu_e}\sim 10^{-8}$ eV,

(ii) For the second generation take 
$c_L=-c_R=.52$, $M_D=1000$ GeV, $M/f=500$ for the suppression of the muon mass to $m_\mu\approx 100$ MeV,
and $M_R = 10^{15}$ GeV for the suppression of the muon neutrino mass 
to $m_{\nu_\mu}\approx 2\cdot 10^{-3}$~eV,

(iii) For the third generation take 
$c_L=-c_R=.51$, $M_D=2000$ GeV, $M/f=100$ for the suppression of the tau mass to $m_\tau\approx 1.7$ GeV,
and $M_R = 6\cdot 10^{14}$ GeV for the suppression of the muon neutrino mass to 
$m_{\nu_\tau}\approx 3\cdot 10^{-2}$ eV.

\begin{figure}
\centerline{
\includegraphics[width=0.3\hsize]{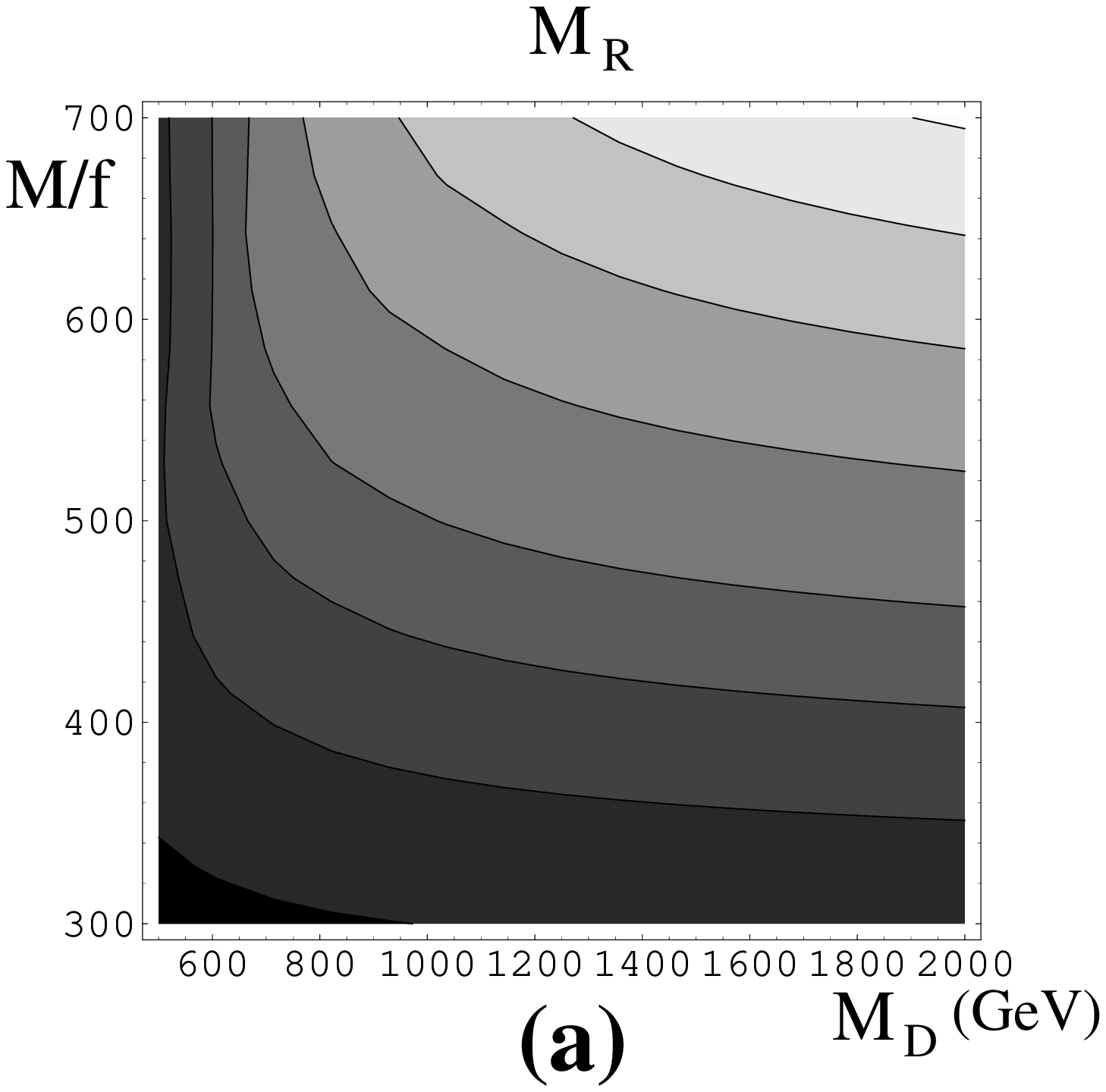}
\hspace{2cm}
\includegraphics[width=0.3\hsize]{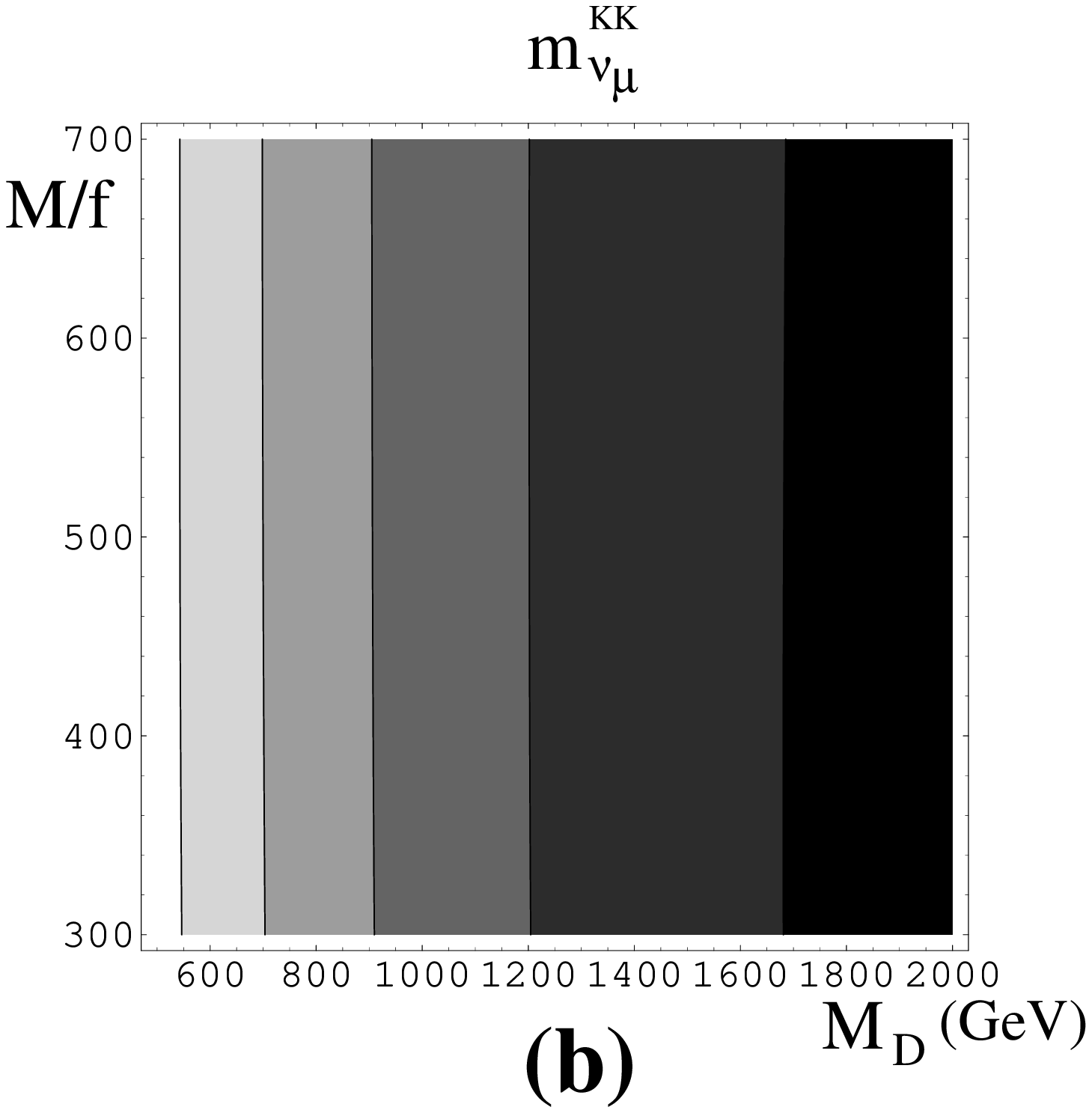}}
\caption{\label{fig:Fit2} {\bf(a)} Contour plot of the value of $M_R$ needed to obtain $m_{\mu}=100$~MeV 
and $m_{\nu_\mu}= 2\cdot 10^{-3}$~eV for
varying values of $M_D$ and $M/f$. The regions starting with the darkest moving toward the 
lighter ones correspond to $M_R = 4,6,8,10,13,16,19,22\cdot 10^{14}$~GeV.
{\bf (b)} Contour plot of the mass of the  first KK excitation of the muon neutrino, keeping fixed $m_{\mu}=100$~MeV 
and $m_{\nu_\mu}= 2\cdot 10^{-3}$~eV and 
varying  $M_D$ and $M/f$. The regions starting with the darkest moving toward the 
lighter ones correspond to $m^{KK}_{\nu_{\mu}} = 300,400,500,600,700$~GeV.
}
\end{figure}

On Fig.~\ref{fig:Fit2}(a) we have plotted the values of the see-saw scale, $M_R$, needed
to reproduce $m_{\mu}=100$~MeV 
and $m_{\nu_\mu}= 2\cdot 10^{-3}$~eV, while varying the other free parameters,
$M_D$ and $M/f$ (we have further imposed that $c_L=-c_R$).
On Fig.~\ref{fig:Fit2}(b), we have also computed the the mass of the first KK excitation
of the muon neutrinos. Its mass decreases as $M_D$ increases and we note that
it is also almost independent of the value of $M/f$ needed to fit the muon as long as we keep
the mass of the muon neutrino fixed.

Another possibility that we will not explore here, due to its relative
complexity, is 
that the charged lepton and neutrino come from two different SU(2)$_R$ doublets as in ref \cite{ADMS}.

\section{Conclusions} 
\setcounter{equation}{0}  
 
We have considered theories with fermions on an extra dimensional interval. We derived the 
consistent BC's from the variational principle and explained how to associate the various BC's 
to different physical situations. We have applied our results to higgsless models of electroweak 
symmetry breaking, and showed that realistic fermion mass spectra can be generated without the 
presence of a Higgs fields both in flat and in warped space.

\section*{Acknowledgments} 
\setcounter{equation}{0} 
We thank Andy Cohen, Nima Arkani-Hamed, Tanmoy Bhattacharya, Alex Friedland,
Markus Luty, Take\-michi Okui, Luigi Pilo, Geraldine Servant, Raman Sundrum for useful conversations. 
The research of C.C. and J.H.
is supported in part by the DOE OJI grant DE-FG02-01ER41206 and in part
by the NSF grants PHY-0139738 and PHY-0098631. C.C. and J.H. thank the T-8 group at the Los Alamos
Laboratory for their hospitality during this work. C.C. and C.G. thank the KITP at UC Santa Barbara
for its hospitality during the course of this work. 
J.T. and Y.S. are supported by the US Department of Energy under contract
W-7405-ENG-36. Y.S. is a Richard P. Feynman fellow at Los Alamos
National Laboratory.
C.G. is supported in part by the RTN European Program
HPRN-CT-2000-00148
and the ACI Jeunes Chercheurs 2068.
J.T. thanks the particle theory group at Boston University for their hospitality
while this work was completed.


\section*{Appendix}
\setcounter{section}{0} 
\renewcommand{\thesection}{\Alph{section}}

\section{Spinor and Gamma Matrix Conventions} 
\label{app:conventions} 
\renewcommand{\theequation}{A.\arabic{equation}} 
\setcounter{equation}{0} 
 
For completeness, we give in this appendix the convention about spinors and Dirac matrices used
throughout the paper. We have mainly followed the conventions of Wess and Bagger~\cite{WB}.

We are working with a mostly ``$-$" space-time signature, $(+----)$, and we have chosen 
the chiral representation of the Dirac gamma matrices:
\begin{equation}
\Gamma^\mu = \left( 
\begin{array}{cc}
0 & \sigma^\mu \\
\bar{\sigma}^\mu & 0 
\end{array}
\right), \mu=0,1,2,3
\ \ \mathrm{and} \ \
\Gamma^5 = \left( 
\begin{array}{cc}
i \mathbf{1}_2 & 0 \\
0 & -i \mathbf{1}_2 
\end{array}
\right)
\end{equation}
where $\sigma^\mu$ and $\bar{\sigma}^\mu$ are the usual Pauli matrices
\begin{eqnarray}
\sigma^0 = - \mathbf{1}_2, \,
\sigma^1 = \left( \begin{array}{cc}
0 & 1 \\
1 & 0 
\end{array}
\right),
\sigma^2 = \left(  \begin{array}{cc}
0 & -i \\
i & 0 
\end{array}
\right),
\sigma^3 = \left( \begin{array}{cc}
1 & 0 \\
0 & -1 
\end{array}
\right),
\\
\bar{\sigma}^0 = - \mathbf{1}_2, \,
\bar{\sigma}^1 =\left(  \begin{array}{cc}
0 & 1 \\
1 & 0 
\end{array}
\right),
\bar{\sigma}^2 = \left(  \begin{array}{cc}
0 & -i \\
i & 0 
\end{array}
\right),
\bar{\sigma}^3 = \left(  \begin{array}{cc}
1 & 0 \\
0 & -1 
\end{array}
\right).
\end{eqnarray}
A famous relation about the Pauli matrices is
\begin{equation}
\sigma^\mu \bar{\sigma}^\nu + \sigma^\nu \bar{\sigma}^\mu
= 2 \eta^{\mu \nu}
\ \ \mathrm{and} \ \
\bar{\sigma}^\mu \sigma^\nu + \bar{\sigma}^\nu \sigma^\mu
= 2 \eta^{\mu \nu}.
\end{equation}

A 5D Dirac spinor is written in terms of a pair of two component spinors
\begin{equation} 
\Psi = \left( 
\begin{array}{c} \chi_\alpha 
\\ 
\tv{13}
\bar{\psi}^{\dot{\alpha}} 
\end{array} 
\right) 
\end{equation} 
The dotted and undotted indices of a two component spinor are raised and lower
with the $2 \times 2$ antisymmetric tensors $\epsilon_{\alpha \beta} = i \sigma^2_{\alpha \beta}$
and $\epsilon_{\dot{\alpha} \dot{\beta}} = i \sigma^2_{\dot{\alpha} \dot{\beta}}$ and their inverse
$\epsilon^{\alpha \beta}=-i \sigma^2_{\alpha \beta}$,
$\epsilon^{\dot{\alpha} \dot{\beta}} = -i \sigma^2_{\dot{\alpha} \dot{\beta}}$ :
\begin{equation}
\chi^\alpha = \epsilon^{\alpha \beta} \chi_\beta
\ \ \mathrm{and}\ \
\bar{\psi}_{\dot{\alpha}} = \epsilon_{\dot{\alpha} \dot{\beta}} \bar{\psi}^{\dot{\beta}}.
\end{equation}
Note also the adjoint relation:
\begin{equation}
(\chi^\dagger)^\alpha = \bar{\chi}^{\dot{\alpha}}.
\end{equation}
Finally, $\chi \psi$ and $\bar{\chi} \bar{\psi}$ denote the two Lorentz invariant scalars:
\begin{equation}
\chi \psi = \chi^\alpha \psi_\alpha
\ \ \mathrm{and}\ \
\bar{\chi} \bar{\psi} = \bar{\chi}_{\dot{\alpha}} \bar{\psi}^{\dot{\alpha}}.
\end{equation}
These products are symmetric
\begin{equation}
\chi \psi = \psi \chi
\ \ \mathrm{and}\ \
\bar{\chi} \bar{\psi} = \bar{\psi} \bar{\chi} .
\end{equation}
%

\section{Examples of KK Decomposition in Flat Space with the Simplest BC's} 
\label{app:KKdecomposition} 
\renewcommand{\theequation}{B.\arabic{equation}} 
\setcounter{equation}{0} 

We present in this appendix explicit examples of KK decomposition of fermions in flat space.

We begin by giving the full KK decomposition in the case of the simplest Dirichlet--Neumann BC's:
there are a priori four different cases to discuss, but the case $\chi_{|0}=\chi_{|L}=0$ is similar to $\psi_{|0}=\psi_{|L}=0$ and the case $\chi_{|0}=\psi_{|L}=0$ is similar to $\psi_{|0} = 
\chi_{|L}=0$. So let us mention the KK decompositions for the latter cases only:

({\it i}) when $\psi_{|0}=\psi_{|L}=0$: there is a zero mode for $\chi$ only, with an exponential 
wave function localized either on the $0$ or $L$ brane depending on the sign of bulk Dirac mass, 
as well as a tower of massive modes mixed between $\chi$ and $\psi$: 
\begin{eqnarray} 
\chi &= & A_0 e^{-my} \xi_0 + \sum_{n=1}^\infty A_n ( \cos k_n  y - \frac{m}{k_n} \sin k_n y)\ \chi_n, \\ 
\psi & = &  -\sum_{n=1}^\infty \frac{m_n}{k_n} A_n \sin k_n y\ \psi_n, 
\end{eqnarray} 
with KK masses $m_n$ solution of the quantization equation:
\begin{equation}
        \label{eq:DN1} 
\sin k_nL=0, \ \ \ \ Ê k_n^2 = m_n^2 - m^2. 
\end{equation}

({\it ii})  when $\psi_{|0}=\chi_{|L}=0$: there is no zero mode at all and the tower of massive modes is given by 
\begin{eqnarray} 
\chi &= &  \sum_{n=1}^\infty A_n ( \cos k_n  y - \frac{m}{k_n} \sin k_n y)\ \chi_n, \\ 
\psi & = &  -\sum_{n=1}^\infty \frac{m_n}{k_n} A_n \sin k_n y\ \psi_n, 
\end{eqnarray} 
with KK masses $m_n$ now solution of the equation:
\begin{equation}
        \label{eq:DN2} 
k_n=m \tan k_n L,  \ \ \ \ Ê k_n^2 = m_n^2 - m^2.
\end{equation}

Let us now discuss the KK decomposition when we impose BC's of the form~(\ref{eq:angleBC}):
\begin{equation} 
s_{0,L}\, \psi_{|0,L} + c_{0,L}\, \chi_{|0,L}  =  0 
\end{equation} 
where $s_{0,L}$  ($c_{0,L}$) stand for the sine (cosine) of some (possibly complex) angles, $\alpha_{0,L}$, that determine which linear combination of the fields on the two boundaries are vanishing. the first thing to notice is that these BC's can be imposed only when the 5D fermion
belongs to a real representation of the gauge group. In this case, the KK modes will be 4D Majorana fermions and the KK decomposition will be of the form
\begin{eqnarray} 
\chi  = \sum_n g_n(y)\, \xi_{n} (x), \\ 
\bar{\psi} = \sum_n f_n(y)\, \bar{\xi}_n (x). 
\end{eqnarray} 
with the spinors $\xi_n(x)$ satisfying the Majorana equation: 
\begin{eqnarray} 
-i \bar{\sigma}^{\mu} \partial_\mu \xi_{n} + m_n\, \bar{\xi}_n = 0, \\ 
-i \sigma^{\mu} \partial_\mu \bar{\xi}_n + m_n^*\, \xi_{n} = 0. 
\end{eqnarray} 
The wave functions are of the form (\ref{eq:wv1})-(\ref{eq:wv2}) with $m_n$ being now replaced by
$|m_n|$. In general there is no zero mode except if there exists a tuning between the
angles $\alpha_{0,L}$ and the bulk mass $m$:
\begin{equation} 
\sin (\alpha_0 - \alpha_L) - \sin (\alpha_0 + \alpha_L) \tanh mL = 0. 
\end{equation} 
For the massive KK modes, as before, the bulk equations give $A_n$ and $B_n$ in terms of $C_n$ and $D_n$ and the boundary equations 
reduce to two complex equations for two complex unknowns. The masses are thus obtained as the roots of a 4 by 4 determinant. After some algebra we obtain  the quantization equation: 
\begin{eqnarray} 
\left( | s_0 s_L |^2 + | c_0 c_L |^2\right) | m_n ^2| \sin^2  k_n L 
-  
| s_0 c_L |^2 (m\, \sin\ k_n L - k_n\,  \cos k_n L)^2 
\nonumber\\ 
- | c_0 s_L |^2  (m\, \sin k_n L + k_n\,  \cos k_n L)^2 
+ ( c_0 s_0^* s_L c_L^* + s_0 c_0^* c_L s_L^*)\, k_n^2 =0. 
\end{eqnarray} 
When all the boundary angles are real, the above equation factorizes into the two simpler 
equations: 
\begin{equation} 
\pm m_n \cos (\alpha_0 - \alpha_L) \, \sin k_n L 
- k_n \sin  (\alpha_0 - \alpha_L) \,  \cos k_n L 
+ m \sin (\alpha_0 + \alpha_L) \, \sin  k_n L 
=0. 
\end{equation} 
In particular, we recover the results (\ref{eq:DN1})--(\ref{eq:DN2}) for the particular angles 
$\alpha_0,\alpha_L=0,\pi/2$.

\section{Boundary Conditions in the Presence of Various Localized Mass/Mixing/Kinetic Terms} 
\label{app:otherBCs} 
\renewcommand{\theequation}{C.\arabic{equation}} 
\setcounter{equation}{0} 

In this appendix, we would like to extend the discussion of
Section~\ref{sec:dynbc}, and present the BC's in the presence of various localized operators
on the boundaries. Of particular
interest are localized terms for fermions in complex representations.
Majorana masses for such theories are forbidden,
however, localized kinetic terms are not, neither are
localized Dirac masses.  We will also discuss examples where bulk fermions are mixing with fermions that are localized at the boundaries.

We will be following the steps outlined at the beginning of Section \ref{sec:dynbc}: we first add the 
localized terms at $y=\epsilon$ to the bulk equations of motion in order to efficiently deal with the 
discontinuities in the wave functions. At the real boundary $y=0$ we impose as always the simplest BC's
\begin{equation}
        \label{eq:DNapp}
\partial_5 \chi_{|0} = 0, \ \
\psi_{|0} = 0, \ \
\partial_5 \chi_{|L} = 0, \ \
\psi_{|L} = 0.
\end{equation}
We then take $\epsilon\to 0$ and identify the relevant BC's at $y=0^+$ and $y=L^-$.

\subsection{Adding localized kinetic terms}

We add an extra boundary kinetic interaction localized at
$y=\epsilon>0$ to one of the two two-component spinors through the $4D$ boundary action
\begin{equation}
S_{\mathrm{4D}} = - \int d^4 x \, i \kappa\,  \bar{\chi} \bar{\sigma}^\mu
\pd_\mu \chi_{| y= \epsilon}.
\end{equation}
The equations of motion are modified to
\begin{eqnarray}
&\displaystyle
-i \bar{\sigma}^{\mu} \partial_\mu \chi 
\left( 1 + \kappa  \delta(y-\epsilon) \right)  
- \partial_5 \bar{\psi}
+ m \bar{\psi} = 0, 
 \\
&\displaystyle
-i \sigma^{\mu} \partial_\mu \bar{\psi} 
+ \partial_5 \chi 
+ m  \chi = 0.
\end{eqnarray}
Integrating the first equation over the delta function gives a jump
condition for $\psi$ which then implies a jump in the derivative of $\chi$:
\begin{equation}
\left[ \bar{\psi} \right]_{|\epsilon} 
= 
- i \kappa \bar{\sigma}^\mu \partial_\mu \bar{\chi}_{|\epsilon}
\ \ \mathrm{and}\ \
\left[ \partial_5 \chi \right]_{|\epsilon} 
= 
 i {\sigma}^\mu \partial_\mu  \left[ \bar{\psi} \right]_{|\epsilon}.
\end{equation}
Using the values~(\ref{eq:DNapp}) of the fields on the boundary, we finally obtain
the BC's we were after
\begin{equation}
       \label{eq:kineticBC}
\bar{\psi}_{|0^+} 
= 
- i \kappa \bar{\sigma}^\mu \partial_\mu \bar{\chi}_{|0^+} 
\ \ \mathrm{and}\ \
\partial_5 \chi_{|0^+}  
= 
 i {\sigma}^\mu \partial_\mu  \bar{\psi}_{|0^+} .
\end{equation}
This is exactly a boundary condition of the form   
$(\psi - i c \sigma^\mu \partial_\mu \bar{\chi})_{|0^+} = 0$ 
obtained from the variational principle in (\ref{bc2}), where 
the parameter $c$ being identified as the coefficient of the boundary localized kinetic term.

\subsection{Adding a Dirac mixing of bulk fermions on the boundary}
\label{sec:MD}

We consider two bulk fermions that are mixed together through a  Dirac mass on the boundary
\beq
        \label{eq:DiracMix}
S_{4D} = \int d^4 x \,M_DL  
\left( \vphantom{\frac{1}{2}}
\psi_2 \chi_{ 1} + \bar{\chi}_1 \bar{\psi}_2 +
\psi_1 \chi_{2} + \bar{\chi}_2 \bar{\psi}_1
 \right)_{|0}
 \eeq
We further need to specify the values of the fields at $y=0$ and $y=L$. We are assuming that as usual 
\begin{equation}
        \label{eq:DNappMD}
\partial_5 {\chi_1}_{|0,L} = 0, \ \
{\psi_1}_{|0,L} = 0, \ \
\partial_5 {\psi_2}_{|0,L} = 0, \ \
{\chi_2}_{|0,L} = 0,
\end{equation}
in such a way that when $M_D\rightarrow0$ there are two zero modes corresponding to $\chi_1$ and
$\psi_2$.

In the present case, there is an ambiguity when we want to push
the interaction mass at a distance $\epsilon$ away from the boundary. Indeed if we were to push
the whole expression~(\ref{eq:DiracMix}), we will end up with discontinuity in the wave functions
of both $\psi_1$ and $\psi_2$ and $\chi_1$ and $\chi_2$, which then requires a regularization of the products like
$\psi_2 \delta (y-\epsilon)$, for instance by an averaging of $\psi_2$ over its limits from
both sides. In this case, using the values~(\ref{eq:DNappMD}) of the fields at $y=0$, we
get the following BC's at $y=0^+$
\begin{eqnarray}
&\displaystyle
{\psi_1}_{|0^+} 
= 
 \frac{M_D L}{1+ \frac{1}{4} M_D^2 L^2}\,
{\psi_2}_{|0^+},
\\
&\displaystyle
{\chi_2}_{|0^+} 
=
- \frac{M_D L}{1+ \frac{1}{4} M_D^2 L^2}\,
{\chi_1}_{|0^+}.
\end{eqnarray}
Another regularization would consist for instance in smoothing the
delta function by a square potential, $M_D / a \Theta (y) \Theta (a-y)$, and then take
the limit $a\rightarrow 0$.  In this case, we would arrive at slightly modified BC's of the form
\begin{eqnarray}
&\displaystyle
{\psi_1}_{|0^+} = 2 \tanh{M_D L/2}\ {\psi_2}_{|0^+},
\\
&\displaystyle
{\chi_2}_{|0^+} = -2 \tanh{M_D L/2}\ {\chi_1}_{|0^+}.
\end{eqnarray}

Another possibility will be to impose the values~(\ref{eq:DNappMD}) on the boundary
before pushing it $\epsilon$ away from the brane. In that case, the
bulk equations of motion will be compatible 
with continuous wave functions for $\psi_2$ and $\chi_1$ and would require
discontinuities in $\psi_1$ and $\chi_2$ only. No further regularization of the delta function
would then be necessary. And the BC's would simply read
\begin{eqnarray}
&\displaystyle
{\psi_1}_{|0^+} =  M_D L\ {\psi_2}_{|0^+},
\\
&\displaystyle
{\chi_2}_{|0^+} = - M_D L\ {\chi_1}_{|0^+}.
\end{eqnarray}

We note that whichever way we go, the BC's are all the same in the limit of small $M_D L$.
And, in the flat case, the lowest eigenmode is a Dirac fermion with mass $M_D$. 
More importantly, the form of the boundary condition is independent of
the regularization of the delta function, and at most the
interpretation of the physical meaning of $M_D$ might depend on it.

\subsection{Mixing with brane  fermions}
\label{sec:BraneFermions}

We now analyze a more general case where, in addition to the bulk
fermions, there are
localized spinors on the boundaries which mix 
with bulk fermions through Dirac  mass terms.  The
addition of a Dirac mixing mass on the $y=L$ brane leads to a 4D action given by
\begin{equation}
        \label{eq:braneferm}
S_{4D}
=  
\int d^4 x  
\left( \vphantom{\frac{1}{2}}
- i \bar{\xi} \bar{\sigma}^\mu \partial_\mu \xi  
- i \eta \sigma^{\mu} \partial_\mu \bar{\eta}
+ ML^{1/2} \left(\eta \chi + \bar{\chi} \bar{\eta} \right)
+ f \left( \eta \xi  + \bar{\xi} \bar{\eta} \right)
\right)_{|y=L},
\end{equation}
where $M$ and $f$ have mass dimension $1$.

To find the corresponding boundary conditions, we follow our by now usual procedure and push the
interactions a small distance $\epsilon$ away from the boundary at $y=L$,
and solve the resulting equations of motion which are  given by
\begin{eqnarray}
&\displaystyle
-i \bar{\sigma}^{\mu} \partial_\mu \chi 
- \partial_5 \bar{\psi} 
+ m \bar{\psi} 
+  ML^{1/2}\, \bar{\eta}\, \delta(y - L + \epsilon) = 0,
\\
& \displaystyle
 -i \sigma^{\mu} \partial_\mu \bar{\psi} 
 + \partial_5 \chi 
 + m \chi = 0,
 \\
& \displaystyle
-i \sigma^{\mu} \partial_\mu \bar{\eta} 
+ ML^{1/2} \chi 
+ f \xi = 0,
\\
 & \displaystyle
-i \bar{\sigma}^{\mu} \partial_\mu \xi 
+ f  \eta = 0. 
\end{eqnarray}
Integrating the first equation over the delta function gives the
jump for $\psi$:
\begin{equation}
\left[ \bar{\psi} \right]_{|L-\epsilon} =  ML^{1/2}\, \bar{\eta},
\end{equation}
which, using the defined value~(\ref{eq:DNapp}) of $\psi$ at $y=L$
implies the following BC at $y=L^-$
\begin{equation}
\bar{\psi}_{|L^-} = - ML^{1/2} \bar{\eta}.
\end{equation}
Then the last two equations can then be combined in the following way:
\begin{equation}
\left( \partial^\mu \partial_\mu + f^2 \right) \bar{\eta} 
= 
- i ML^{1/2}\, \bar{\sigma}^{\mu}
\partial_\mu \chi_{|L}.
\end{equation}
Taking the limit  $\epsilon \rightarrow0$, we finally arrive at the BC's:
\begin{eqnarray}
& \displaystyle
\bar{\psi}_{|L^-} 
=
- M L^{1/2} \bar{\eta},
\\
        \label{eq:DiracBraneBC}
& \displaystyle
(\partial^\mu \partial_\mu +f^2) \bar{\psi}_{|L^-} 
=
i  M^2L \bar{\sigma}^{\mu}
\partial_\mu \chi_{|L^-}.
\end{eqnarray}

We notice that interactions (\ref{eq:braneferm}) lead to a
generalization of the boundary condition obtained by adding a brane
localized kinetic term. Indeed, in the limit when $f$ and $M$ tend to
infinity with a fixed ratio, the boundary condition (\ref{eq:DiracBraneBC})
reduces precisely to (\ref{eq:kineticBC}). This is
expected since in the limit of large mass $f$ of the brane fermions
they must be integrated out. This leaves one massless fermion, which
is a linear combination of $\xi$ and $\chi$, and its kinetic term contains a delta-function
contribution, originating from the localized kinetic term for $\xi$, which is proportional to
$M^2/f^2$. When we constructed a realistic pattern for the quark and lepton masses
in Section~\ref{sec:WarpedNoHiggs}, we explicitly introduced brane localized fermions
but it should be kept in mind that a similar spectrum can be obtained just by introducing
on the Planck brane localized kinetic terms for the bulk fermions.

When the bulk fermion appears to be neutral with respect
to the residual gauge symmetry on the boundary, it can couple to
a Majorana fermion on the brane. So let us see how the previous BC's are modified in that case.
The 4D boundary localized action is
\begin{equation}
        \label{eq:MajoranaBrane}
S_{4D}
= 
\int d^4 x 
\left( \vphantom{\frac{1}{2}}
-i \eta \sigma^{\mu} \partial_\mu \bar{\eta}
+ M L^{1/2} \left(\eta \chi + \bar{\chi} \bar{\eta} \right)
+ \sfrac{1}{2} f(\eta\eta +\bar\eta\bar\eta)
\right)_{|y=L}
\end{equation}
where $M$ and $f$ have again a mass dimension $1$.

The same procedure of pushing the interaction $\epsilon$ away from the boundary and taking the limit 
$\epsilon \to 0$ leads to a jump in $\psi$ and a jump in $\partial_5 \chi$ and the BC's finally read:
\begin{eqnarray}
        \label{eq:MajoranaBraneBC}
& \displaystyle
\bar{\psi}_{|L^-}
=
- M L^{1/2} \bar{\eta} ,
\\
& \displaystyle
\label{eq:majbrnbc2}
\partial_5 \chi_{|L^-} 
= 
(-M^2L \, \chi + f \psi)_{|L^-}.
\end{eqnarray}
%



\begin{thebibliography}{99} 

\bibitem{CGMPT}
C.~Cs\'aki, C.~Grojean, H.~Murayama, L.~Pilo and J.~Terning,
{\tt hep-ph/0305237}.

\bibitem{otherunitarity}
R.~S.~Chivukula, D.~A.~Dicus and H.~J.~He,
Phys.\ Lett.\ B {\bf 525}, 175 (2002)
[{\tt hep-ph/0111016}];
R.~S.~Chivukula, D.~A.~Dicus, H.~J.~He and S.~Nandi,
{\tt hep-ph/0302263};
R.~S.~Chivukula and H.~J.~He,
Phys.\ Lett.\ B {\bf 532}, 121 (2002)
[{\tt hep-ph/0201164}];
S.~De Curtis, D.~Dominici and J.~R.~Pelaez,
Phys.\ Lett.\ B {\bf 554}, 164 (2003)
[{\tt hep-ph/0211353}];
and Phys.\ Rev.\ D {\bf 67}, 076010 (2003)
[{\tt hep-ph/0301059}];
Y.~Abe, N.~Haba, Y.~Higashide, K.~Kobayashi and M.~Matsunaga,
{\tt hep-th/0302115}.

\bibitem{SonStephanov}
D.~T.~Son and M.~A.~Stephanov,
{\tt hep-ph/0304182}.

\bibitem{A5Higgs}
C.~Cs\'aki, C.~Grojean and H.~Murayama,
Phys.\ Rev.\ D {\bf 67}, 085012 (2003)
[{\tt hep-ph/0210133}];
I.~Gogoladze, Y.~Mimura and S.~Nandi,
Phys.\ Lett.\ B {\bf 560}, 204 (2003) {\tt [hep-ph/0301014]};
Phys.\ Lett.\ B {\bf 562}, 307 (2003)
{\tt [hep-ph/0302176]};
C.~A.~Scrucca, M.~Serone and L.~Silvestrini,
{\tt hep-ph/0304220}.

\bibitem{ST}
M.~E.~Shaposhnikov and P.~Tinyakov,
Phys.\ Lett.\ B {\bf 515}, 442 (2001) {\tt [hep-th/0102161]}.

\bibitem{CGPT}
C.~Cs\'aki, C.~Grojean, L.~Pilo and J.~Terning,
{\tt hep-ph/0308038}.


\bibitem{nomura}
Y.~Nomura,
{\tt hep-ph/0309189}.

\bibitem{RS}
L.~Randall and R.~Sundrum, 
Phys.\ Rev.\ Lett.\  {\bf 83}, 4690 (1999)
{\tt [hep-th/9906064]};
Phys.\ Rev.\ Lett.\  {\bf 83}, 3370 (1999)
{\tt [hep-ph/9905221]}.


\bibitem{RSbulk}
H.~Davoudiasl, J.~L.~Hewett and T.~G.~Rizzo,
Phys.\ Lett.\ B {\bf 473}, 43 (2000)
{\tt [hep-ph/9911262]};
Phys.\ Rev.\ D {\bf 63}, 075004 (2001) {\tt [hep-ph/0006041]};
A.~Pomarol,
Phys.\ Lett.\ B {\bf 486}, 153 (2000)
{\tt [hep-ph/9911294]};
Phys.\ Rev.\ Lett.\  {\bf 85}, 4004 (2000)
{\tt [hep-ph/0005293]};
S.~Chang, J.~Hisano, H.~Nakano, N.~Okada and M.~Yamaguchi,
Phys.\ Rev.\ D {\bf 62}, 084025 (2000)
{\tt [hep-ph/9912498]};
S.~J.~Huber and Q.~Shafi,
Phys.\ Rev.\ D {\bf 63}, 045010 (2001)
{\tt [hep-ph/0005286]};
L.~Randall and M.~D.~Schwartz,
Phys.\ Rev.\ Lett.\  {\bf 88}, 081801 (2002)
{\tt [hep-th/0108115]};
JHEP {\bf 0111}, 003 (2001)
{\tt [hep-th/0108114]};
S.~J.~Huber, C.~A.~Lee and Q.~Shafi,
Phys.\ Lett.\ B {\bf 531}, 112 (2002)
{\tt [hep-ph/0111465]};
J.~L.~Hewett, F.~J.~Petriello and T.~G.~Rizzo,
JHEP {\bf 0209}, 030 (2002)
{\tt [hep-ph/0203091]};
H.~Davoudiasl, J.~L.~Hewett and T.~G.~Rizzo,
{\tt hep-ph/0212279};
M.~Carena, E.~Ponton, T.~M.~Tait and C.~E.~Wagner,
Phys.\ Rev.\ D {\bf 67}, 096006 (2003)
{\tt [hep-ph/0212307]};
M.~Carena, A.~Delgado, E.~Ponton, T.~M.~Tait and C.~E.~Wagner,
{\tt hep-ph/0305188}.

\bibitem{CET}
C.~Cs\'aki, J.~Erlich and J.~Terning, Phys.\ Rev.\ D {\bf 66}, 064021
(2002)
{\tt [hep-ph/0203034]}.

\bibitem{ADMS}
K.~Agashe, A.~Delgado, M.~J.~May and R.~Sundrum,
JHEP {\bf 0308}, 050 (2003)
{\tt [hep-ph/0308036]}.




\bibitem{holography}
N.~Arkani-Hamed, M.~Porrati and L.~Randall,
JHEP {\bf 0108}, 017 (2001)
{\tt [hep-th/0012148]};
R.~Rattazzi and A.~Zaffaroni,
JHEP {\bf 0104}, 021 (2001)
{\tt [hep-th/0012248]};



\bibitem{BPR}
R.~Barbieri, A.~Pomarol and R.~Rattazzi,
{\tt hep-ph/0310285}.



\bibitem{matthias} 
Y.~Grossman and M.~Neubert, 
Phys.\ Lett.\ B {\bf 474}, 361 (2000) 
{\tt [hep-ph/9912408]}. 

\bibitem{GherPom}
T.~Gherghetta and A.~Pomarol,
Nucl.\ Phys.\ B {\bf 586}, 141 (2000) {\tt [hep-ph/0003129]};


\bibitem{kaptait} 
D.~E.~Kaplan and T.~M.~Tait, 
JHEP {\bf 0111}, 051 (2001) 
{\tt [hep-ph/0110126]}. 

\bibitem{HS1}
S.~J.~Huber and Q.~Shafi,
Phys.\ Lett.\ B {\bf 498}, 256 (2001)
{\tt [hep-ph/0010195]};
Phys.\ Lett.\ B {\bf 512}, 365 (2001)
{\tt [hep-ph/0104293]};
Phys.\ Lett.\ B {\bf 544}, 295 (2002)
{\tt [hep-ph/0205327]}.

\bibitem{nomurasmith}
Y.~Nomura and D.~R.~Smith,
Phys.\ Rev.\ D {\bf 68}, 075003 (2003)
{\tt [hep-ph/0305214]}.

\bibitem{HSrecent}
S.~J.~Huber and Q.~Shafi,
{\tt hep-ph/0309252}.

\bibitem{WB}
J.~Wess and J.~Bagger,
``Supersymmetry And Supergravity,''
Princeton, USA: Univ. Pr. (1992) 259 p. ;
%
J.~A.~Bagger,
{\tt hep-ph/9604232}.

\bibitem{Kaplan}
D.~B.~Kaplan,
Phys.\ Lett.\ B {\bf 288}, 342 (1992)
{\tt [hep-lat/9206013]}.

\bibitem{BFZ}
J.~A.~Bagger, F.~Feruglio and F.~Zwirner,
Phys.\ Rev.\ Lett.\  {\bf 88}, 101601 (2002)
{\tt [hep-th/0107128]};


\bibitem{otherBagger}
J.~Bagger, F.~Feruglio and F.~Zwirner,
JHEP {\bf 0202}, 010 (2002)
{\tt [hep-th/0108010]};
J.~Bagger and M.~Redi,
{\tt hep-th/0310086}.





\bibitem{walking}B. Holdom,  Phys. Rev. {\bf D24} 1441  (1981);
B. Holdom, Phys. Lett. {\bf B150} 301 (1985);
K. Yamawaki, M. Bando, and K. Matumoto, Phys. Rev. Lett.
{\bf 56} 1335  (1986);
T. Appelquist, D. Karabali, and L.C.R. Wijewardhana, Phys. Rev.
Lett. {\em 57} 957  (1986);
T. Appelquist and L.C.R. Wijewardhana, Phys. Rev. {\bf D35} 774  (1987);
T. Appelquist and L.C.R. Wijewardhana, Phys. Rev. {\bf D36} 568 (1987).



\bibitem{Burdman}
G.~Burdman,
{\tt hep-ph/0310144}.

\bibitem{Hiller}
G.~Hiller and M.~Schmaltz,
Phys.\ Lett.\ B {\bf 514}, 263 (2001)
{\tt hep-ph/0105254}.
G.~Hiller and M.~Schmaltz,
Phys.\ Rev.\ D {\bf 65}, 096009 (2002)
{\tt hep-ph/0201251}.



\end{thebibliography}
 \end{document}